\documentclass[sigconf]{acmart}

\usepackage{graphicx}
\usepackage{mathrsfs}

\usepackage{lipsum}
\usepackage[english]{babel}
\usepackage{comment}

\newcommand{\yuzhe}[1]{\textcolor{orange}{yz: #1}}

\newcommand{\g}{\ensuremath{\mathsf{g}}}

\usepackage{fancyhdr} 
\usepackage{float}

\usepackage{autobreak} 
\usepackage{wrapfig} 

\usepackage{stackengine}
\usepackage{scalerel}
\usepackage{xcolor}
\newcommand\dangersign[1][2ex]{%
  \renewcommand\stacktype{L}%
  \scaleto{\stackon[1.3pt]{\color{red}$\triangle$}{\tiny !}}{#1}%
}
\definecolor{darkgreen}{RGB}{0, 128, 0}

\usepackage{url}

\usepackage{hyperref}
\hypersetup{
    colorlinks=true,
    linkcolor=blue,
    filecolor=magenta,      
    urlcolor=magenta,
    citecolor = magenta,
    pdftitle={An Example},
    pdfpagemode=FullScreen,
    }
\newcommand{\hlhref}[2]{\href{#1}{\textcolor{blue}{\underline{#2}}}}

\usepackage{pifont}
\usepackage{bbding} 
\usepackage{pifont}
\newcommand{\cmark}{\textcolor{green!80!black}{\ding{51}}}

\usepackage{makecell}
\usepackage{colortbl}

\usepackage{subcaption}
\usepackage{tikz} 
\usepackage{comment}
\usepackage{filecontents}
\usepackage{url}
\usepackage{hyperref}
\usepackage{multirow}

\usepackage{algorithmic}
\usepackage{algorithm}
\usepackage{graphicx}
\usepackage{textcomp}
\usepackage{xcolor}
\usepackage{amsthm}
\usepackage{booktabs}
\usepackage[utf8]{inputenc}

\usepackage{dashbox}
\usepackage{todonotes}
\usepackage{enumitem}   
\usetikzlibrary{shapes.callouts}
\usepackage{listings}
\usepackage{trace}
\usepackage{mdframed}

\theoremstyle{plain}
 
\newtheorem{thm}{Theorem}   
\newtheorem{lem}{Lemma}  
\newtheorem{defi}{Definition}
\theoremstyle{definition}

\newtheorem*{prf}{Proof}

\newtheorem{exmp}{Example}
\newtheorem{rmk}{Remark}

\usepackage{threeparttable} 
\tikzset{%
    pics/sema/.style args={#1/#2/#3}{code={%
        \ifstrequal{#2}{0}{%
            \node[circle,minimum width=1mm,draw,fill=#1] {};
        }{%
            \tkzDefPoint(0,0){O}
            \tkzDrawSector[R,fill=#1](O,1mm)(90,90-#2)
            \tkzDrawSector[R,fill=#3](O,1mm)(90-#2,90-360)
    }
    }},
}

\usepackage{tikz}

\usepackage{ulem}

\renewcommand\footnotetextcopyrightpermission[1]{} 
\begin{document}
\title{How to Rationally Select Your Delegatee in PoS}

\author{Yuzhe Zhang, Qin Wang, Shiping Chen, Chen Wang}
\affiliation{
\vspace{0.3em}
\textit{CSIRO Data61, Australia} 
}

\begin{abstract}

This paper centers around a simple yet crucial question for everyday users: \textit{How should one choose their delegated validators within proof-of-stake (PoS) protocols, particularly in the context of Ethereum 2.0?} This has been a long-overlooked gap, as existing studies have primarily focused on inter-committee (validator set) behaviors and activities, while neglecting the dynamic formation of committees, especially for individual stakeholders seeking reliable validators. Our study bridges this gap by diving into the delegation process (\textit{normal users delegate their small-value tokens to delegatees who later act as validators}) before entering an actual consensus phase. 

We propose a Bayesian model to quantify normal users' trust in delegatees, which we further incorporate into a game-theoretical model to simulate users' reactions against a set of critical factors identified through extensive research (including 10+ staking service providers as well as 30+ PoS blockchains). Our results reveal that users tend to choose their delegatees and utilize their tokens by carefully weighing the delegation cost, the behaviors of other users, and the reputation of delegatees, ultimately reaching a Nash equilibrium. Unfortunately, the collective trend significantly increases the likelihood of token concentration on a small number of delegatees.


\end{abstract}


\maketitle


\section{Introduction}
\label{sec-intro}

After the Merge~\cite{themerge}, Ethereum officially transitions from the \textit{proof-of-work} (PoW) consensus mechanism to \textit{proof-of-stake} (PoS)~\cite{posetheruem}, marking the advent of its 2.0 version. This upgrade brings forth several enhancements, including reduced entry barriers, improved energy efficiency (99.5\% \cite{themerge}), and more robust crypto-economic incentives (e.g., inspiring DeFi protocols and staking services), catering to a wider spectrum of users. Beyond Ethereum, a multitude of industry-leading blockchains have embraced PoS as their consensus mechanism. These prominent PoS-based blockchain platforms include Cosmos, Polygon, Tezos, BNB Chain, Avalanche, Fantom, Cardano, Solana, Kuasama, Polkadot, Aptos, NEAR, Flow, Secret Network, SUI, Oasis, Kava, Band Protocol, and Casper Network. PoS-based blockchain shares have now increased to over 48\% of the entire cryptocurrency market ($\mathsf{\#CoinMarketCap}$).

We focus our attention specifically on the Ethereum ecosystem due to its representativeness. To date (as of Oct. 2023), Ethereum has become the second-biggest cryptocurrency in terms of market capitalization (US\$195,154,903,316, $\mathsf{\#CoinMarketCap}$) and the largest PoS blockchain platform. Ethereum boasts a significant number of registered validators, with 856,167 validators in total. Additionally, the platform has 150,622 actual depositors ($\mathsf{\#Dune}$\footnote{Data source from $\mathsf{Dune}$: \#1 \url{https://dune.com/conelohan/ethereum-pos-and-merge-sql-challenge-twigblock},  \#2 \url{https://dune.com/hildobby/eth2-staking}. \#3 \url{https://dune.com/21co/ethereum-staking-and-withdrawals}, \#4 \url{https://dune.com/chorus_one_research/ethereum-mev-data}. [Oct. 2023].}). The total staked deposits amount to 27,556,644 ETH (equiv. US\$45B), representing an estimated staking ratio of 22.77\%. This indicates a substantial portion of the Ethereum network's native cryptocurrency is being actively staked by validators.

Validators play a pivotal role in PoS-based Ethereum. A validator assumes responsibility for a wide array of processes throughout the consensus procedures, which includes managing stakes, coordinating committees, proposing blocks, validating neighboring blocks, and casting votes for finalization (as detailed in \underline{Sec.\ref{sec-background}}). Consequently, becoming a validator offers various potential income streams, including staking annual reward rates (typically ranging from 3\% to 9\%), block rewards, transaction fees (with a consensus layer APR of up to 3.18\%), and the ability to extract profits, such as  \textit{miner extractable value} (MEV~\cite{daian2020flash}, resulting in a net profit of US\$672,889). Given these incentives, becoming a validator emerges as an attractive option for rational Ethereum participants.

\smallskip
\noindent\textbf{The forgotten majority.} However, becoming a validator demands a minimum of 32 ETH~\cite{posetheruem} (equiv. US\$52,556), a threshold that can pose a significant barrier to entry for many participants. Based on our calculation, the ratio of active validators (150,622, $\mathsf{\#Dune}$) to the total cumulative users (246.52M, based on the hint of \textit{unique Ethereum addresses}\footnote{$\mathsf{\#YChart}$ \url{https://ycharts.com/indicators/ethereum_cumulative_unique_addresses}.}) stands at a mere 0.61\%. This indicates that only a very small fraction of normal users, even when accounting for users with multiple accounts, can afford the substantial deposit required to become validators. Notably, despite normal users forming the majority of this permissionless network, only a minuscule portion can actively participate in its maintenance. This lack of accessibility raises concerns as it appears that the broader market often overlooks the needs of these users. For many of these participants, the only viable option (more in \underline{Sec.\ref{subsec-validator}}) to participate in the network is by delegating their stakes to eligible validators.

\smallskip
\noindent\textbf{Delagation reliance.} Existing ways of delegation for normal users are either \textit{custodial}\footnote{We skip custodial delegation due to its limited usage for CEXes (\underline{Sec.\ref{subsec-validator}}).} (without private keys) or \textit{non-custodial} (holding keys). Besides depositing stakes within exchanges, users prefer to delegate their stakes to one or multiple reputable existing validators, often referred to as stake service providers (\underline{Sec.\ref{subsec-stake}}). This approach allows users to delegate only their tokens, rather than giving up full control of their accounts.  Consequently, it has given rise to a category of services specializing in collecting small investments from normal users and distributing rewards based on their proportional holdings. These service providers typically charge fees, known as commission fees, as compensation for their services.

\smallskip
\noindent\textbf{\textcolor{red}{{\dangersign}} User's decision?} Then, \textit{how do users select their delegated validators?} This is a fundamental yet crucial question. Assuming a simple case, when a normal user Alice opens the wallet and decides to delegate her assets, she confronts a multitude of options. Should she choose the validator with the lowest service fees, the one with the largest user base, the validator with the most substantial deposits, or perhaps one they are already familiar with? It's a complex decision influenced by various factors. To our surprise, despite the importance of this decision, our extensive research, including a thorough investigation detailed in \underline{Apx.\ref{apx-rw}}\footnote{\textbf{Abbreviation}: definition (Dfn.), equation (Eq.). theorem (Thm.), lemma (Lm.), appendix (Apx.), algorithm (Alg.), table (tab.), figure (Fig.), section (sec.).}, revealed a significant gap in existing research. While a few recent studies \cite{li2023how,li2023liquid} have delved into the delegation process in \textit{permissioned} blockchains like EOS and STEEM, there is a noticeable absence of focus on \textit{permissionless} chains, including Ethereum 2.0. This leads us to explore their delegation patterns, processes, and potential impacts.



\smallskip
\noindent\textbf{Contributions.} We approach these goals through a series of efforts.

\smallskip
\noindent\textbf{\ding{172} We are the first to identify the delegation problem in permissless blockchains} (\underline{Sec.\ref{sec-background}}).  Our motivation originates from our practical experiences participating in Ethereum PoS staking, as well as several other PoS-based projects (e.g., Oasis). We encountered difficulties when deciding which validator was most suitable for us. We embarked on a study from various aspects including PoS-based Ethereum (\underline{Sec.\ref{subsec-ethereum}}), participant behaviors (\underline{Sec.\ref{subsec-validator}}), staking services (\underline{Sec.\ref{subsec-stake}}) and surrounding works  (\underline{Apx.\ref{apx-rw}}). Our systematic studies enable us to develop a better understanding of the delegation process and identify several critical factors that significantly influence users' decisions, such as \textit{provider's reputation}, \textit{staking scale}, \textit{depositor scale}, and \textit{commission fees}. Here, we emphasize \textit{permissionless} because, unlike permissioned blockchains, permissionless chains align more closely with Nakamoto's original idea of decentralization~\cite{nakamoto2008bitcoin}. Additionally, such projects continue to dominate crypto markets (53\%, $\mathsf{CoinMarketCap}$) and boast large communities. 

\smallskip
\noindent\textbf{\ding{173} We develop a suite of models to maximize the simulation of the delegation process, suitable for both \textit{permissioned} and \textit{permissionless} settings} (\underline{Sec.\ref{sec-construction}}). In the context of PoS protocols, a delegation involves both a set of delegators $A$ (equiv. users) and delegatees $V$ (equiv. validators).
Our model is designed to seamlessly accommodate both blockchain configurations without sacrificing its forward compatibility with previous definitions.

\noindent \textcolor{teal}{\ding{46}} \textit{We introduce adjustability to set the validity of both delegators and delegatees, allowing us to simulate the dynamic joining and leaving of participants in permissionless settings.} In line with the common practice of configuring delegation models (e.g., \cite{zhang2021power,li2023liquid,li2023how}), we assume a finite set $V$, which has a fixed size (e.g., $|V|=m$) during an entire epoch for permissioned blockchains where the committee remains static. 
However, our game-theoretical model (i.e., the validator selection game (VSG) defined in \underline{Sec.\ref{subsec-gamedef}}) can potentially capture the dynamic feature of permissionless settings in three ways: 

\noindent\hangindent 1em \textcolor{teal}{$\diamond$} The VSG can be modified by setting that $A$ and $V$ are adjustable inter-epoch under constraint $P=A\cup V$ and adding the utility function of validators.
Then, each participant in the game is able to exchange their role between a user and a validator, and the game becomes an evolutionary game \cite{kim2019mining}.

\noindent\hangindent 1em \textcolor{teal}{$\diamond$}  We can easily modify each user's strategy space ($\Sigma_i\in V\times \mathbb{R}_{\ge 0}$, \underline{Sec.\ref{subsec-gamedef}}) by removing specific validators, such that we render the removed validators silent participants.

\noindent\hangindent 1em \textcolor{teal}{$\diamond$}  We can easily extend this permission adaptability of silent mode to users ($A$) by simply setting a user's budget ($\mathbf{b}$, \underline{Dfn.\ref{defi-vsgame}}) to $0$.



\noindent \textcolor{teal}{\ding{46}}
\textit{We introduce a novel metric, "trust," to emulate delegators' belief in validators' integrity information within the market (\underline{Sec.\ref{subsec-trust}}).} This metric ($T$) is updated within a Bayesian probabilistic framework (see \underline{Fig.\ref{fig:bayes}}), incorporating factors revealing validators' integrity, such as \textit{brand reputation} and \textit{rumors}, and the other users' \textit{judgment} on such integrity factors that contribute to a validator's trustworthiness.

Notably, our proposed trust metric mainly reflects users' belief in validators' {\it intrinsic} motivation to leave the market, as well as other users' judgment on this motivation. In practice, validators might be motivated to leave the market by various {\it extrinsic} factors, e.g., their cost to run a client, their received number of delegation tokens, and so on. Our model is ready to be extended to take such extrinsic factors into consideration and to simulate validators' behavior.




\noindent \textcolor{teal}{\ding{46}} In accordance with this approach, \textit{we have developed a Validator Selection Game (VSG) designed to replicate real-world user delegation scenarios} (\underline{Sec.\ref{subsec-validator}}). 
By formally analyzing VSGs, we gain insight into how delegators compete to secure their utility in delegation scenarios.
Our game (\underline{Dfn.\ref{defi-vsgame}}) takes into consideration all the factors mentioned earlier, including participants ($A$, $V$), user attributes like accuracy and error ($\mathbf{q}$, see \underline{Tab.\ref{tab:notation}}), budget ($\mathbf{b}$), strategy ($\Sigma$), validator characteristics such as integrity ($\mathbf{p}$), and external elements like commission fees ($\mathbf{c}$). This holistic approach covers a wide range of behaviors that users may exhibit throughout their engagement with the game. Furthermore, we have formulated a \textit{utility} function (\underline{Dfn.\ref{defi-utility}}) based on this model, which accounts for all these behaviors.


\smallskip
\noindent\textbf{\ding{174} We both theoretically and practically analyze the dynamic delegation via game theory, elucidating the methods to attain a Nash Equilibrium (NE)} (\underline{Sec.\ref{sec-game}}\&\underline{Sec.\ref{sec-experiment}}). 

\textit{Theoretically}, we investigate the existence and feature of Nash equilibria (NE) in Validator Selection Games (VSGs) under various conditions. This includes scenarios with a single validator (\underline{Sec.\ref{subsec-single}}), with multiple homogeneous validators (\underline{Sec.\ref{subsec-homo}}), as well as with commission-free validators (\underline{Sec.\ref{subsec-cfree}}) in VSGs. Our analyses yield a series of proofs that demonstrate the existence of NE under specific conditions (\underline{Thm.\ref{thm:NEsingle}} to \underline{Thm.\ref{thm:NEcf}}).

\textit{Practically}, we conduct a series of experiments to assess the performance of our game by varying parameter configurations (\underline{Sec.\ref{sec-experiment}}).
In simulations, we design an algorithm modeling that delegators noisily execute an optimization of their utility against the other delegators' behaviors, referred to as the \textit{best response} (Dfn.\ref{defi-bestresponse}). Given the inherent uncertainty in these games, we focus on several key parameters that best elucidate our game. The collective results (\underline{Sec.\ref{subsec-result}}) demonstrate alignment with our theoretical analyses.

\smallskip
\noindent \textcolor{teal}{\ding{171}} \textbf{We further offer several key takeaways from our study.}

\smallskip
\noindent \textcolor{teal}{$\diamond$} A non-misled delegator is more inclined to trust a delegatee selected by a larger number of delegators and possessing a strong reputation. This represents a type of the \textit{80-20 rule}~\cite{koch201180} (a.k.a., Pareto Principle) existing in staking markets proved by our theory.

\noindent \textcolor{teal}{$\diamond$} A rational delegator makes choices regarding their delegatee and the number of tokens to delegate by carefully balancing their trust in the delegatees and the associated delegation costs.

\noindent \textcolor{teal}{$\diamond$} A large amount of tokens might be concentrated on a limited number of delegatees who have a good balance between high reputation and low delegation cost. This trend becomes more obvious if delegators are able to choose their best strategies more accurately, or repeatedly alter their strategies.

\section{Understanding Ethereum 2.0} \label{sec-background}

\subsection{System Overview}\label{subsec-ethereum}
 
\noindent\textbf{Network assumption}. Similar to its previous version, Ethereum 2.0 operates on a \textit{partially synchronous model}~\cite{dwork1988consensus}, ensuring that messages will eventually be delivered, albeit with an unknown but finite upper-bound time delay.

\smallskip
\noindent\textbf{Entities}. Two types of participants are involved: (i) \textit{normal users} (\textcolor{teal}{delegator}s in this work) engage in the blockchain network for simply staking (thus becoming stakeholders) or trading tokens. They typically access the network through lightweight clients, such as web browsers, wallets, or mobile apps. (ii) \textit{validators} (\textcolor{teal}{delegatee}s) are either individual participants or groups representing normal users who delegate their tokens. They take on the responsibility of performing the consensus mechanism. In the context of PoW protocols, validators are referred to as miners. They play a pivotal role in maintaining the safety and liveness of blockchain systems~\cite{garay2015bitcoin}.

\smallskip
\noindent\textbf{Chain operation}. Ethereum 2.0 operates on three key pillars: execution, consensus, and incentive.

\textit{Execution}: Execution focuses on the responsibilities of validating and executing transactions. Notably, a recent shift known as \textit{proposer-builder separation} (PBS)~\cite{heimbach2023ethereum} has emerged, aiming to decouple the tasks of ordering transactions (the builder) from those proposing the block (the proposer). This step streamlines transaction packaging and block production.

\textit{Consensus} (\textit{Casper}~\cite{buterin2020combining}): Ethereum 2.0 utilizes a combination of two fundamental primitives: the fork choice rule \textit{LMD GHOST}~\cite{neu2021ebb} and the finality gadget \textit{Casper FFG}~\cite{buterin2017casper}. LMD GHOST\footnote{LMD GHOST is a variant of the GHOST~\cite{sompolinsky2013accelerating} (Greedy Heaviest-Observed Sub-Tree) rule that is based on each participant's most recent vote (LMD, latest message-driven).}  builds upon the heaviest chain rule akin to Nakamoto consensus~\cite{sompolinsky2013accelerating} while simultaneously considering the latest message from each validator. Casper FFG introduces a gadget capable of adding finality to an underlying consensus protocol through the specification of epochs and checkpoints. This step finalizes block confirmations, resolves forks, and facilitates chain growth.

\textit{Incentive}: It primarily defines the policies~\cite{buterin2020incentives} for rewarding honest validators and imposing penalties on malicious ones~\cite{eth2023cryptoeconomics}.

\subsection{Validator and Normal User}\label{subsec-validator}

\noindent\textbf{Validators becomes more important}. Validators have a range of responsibilities: (i) periodically acting as block proposers to generate blocks after validating and ordering transactions from the mempool; (ii) continuously verifying the validity of blocks created by other validators and attesting them to the canonical chain; (iii) participating in consensus operations and voting for finalization; (iv) taking part in sync committees to ensure the network remains operational; and (v) managing stakes and distributing profits, with half of the rewards going to normal users (individual shareholders).

Compared to miners in PoW, validators play a more crucial role in PoS. PoS requires continuous \textit{active participation} from over two-thirds of the validators to maintain blockchain progress. Validators do not incur the same tangible costs (e.g., electricity expenses in PoW), making it easier for them to accumulate stakes from normal users, thereby increasing their influence within the chain. As evidence, validators exercise control over more than 87\% of the ETH on-chain assets ($\mathsf{\#Dune}$) in practice.

\smallskip
\noindent\textbf{Benefits for validators}. Ethereum validators have the opportunity to earn rewards from various sources. The first line is based on faithful consensus-related activities: validators can engage in the consensus process by proposing blocks (approx. 0.04 ETH per successful proposal~\cite{ben2023upgrading}). They can also attest to blocks, including attesting to the source epoch, target epoch, and head block (0.00001 ETH per attestation). Meanwhile, a small portion of rewards can be contributed by participating sync committee process~\cite{ben2023upgrading}.  The second line is to report misbehaves: validators can report dishonest validators and receive whistle-blowing rewards~\cite{heimbach2023ethereum}. Another line is to chase extra profits: validators can generate MEV profits by running MEV-boost services~\cite{mevboost} (0.1 ETH per block~\cite{benjamin2023eth}). The gross profit by arbitrage reaches US\$2,297,234 as of Oct 2023 ($\mathsf{\#Dune}$).

Correspondingly, validators also face penalties for misconduct, such as producing two blocks for the same slot, which can result in their stakes being slashed (at least 1/32 of their staked ETH~\cite{ben2023slash}).

\smallskip
\noindent\textbf{Barriers in becoming validators}. In Ethereum, becoming a validator to take responsibility for the entire network and earn revenues is required to deposit a minimum of 32 ETH, which is worth over US\$53k (based on the price as of Oct 1st, 2023). This high barrier poses a significant challenge for ordinary users with limited cryptoassets. Our investigation reveals that only 0.61\% of Ethereum addresses hold more than 32 ETH (equiv. US\$52,556) in their accounts, which has also been mentioned in \underline{Sec.\ref{sec-intro}}.

\smallskip
\noindent\textbf{How do users participate?} Normal users join in the game by three ways (i) becoming a validator by depositing 32 ETH and maintaining a full node; (ii) delegating their tokens via staking service providers, where users only send tokens to the delegatee's address while still holding their account private keys (\textit{non-custodial}); and (iii) delegating their tokens within platforms provided by service providers like centralized exchanges (e.g., Binance, Coinbase) whereas the private keys are owned by these providers. Our research is centered around the second route.

\subsection{Staking Services}\label{subsec-stake}

Following above, there are several ways of staking~\cite{howtostake}: \textit{solo-home staking}, delegating tokens to a \textit{staking as a service} provider, \textit{pooled staking} and creating accounts in \textit{centralized exchanges} (CEXes).

We investigate two major types of such staking services in the market: \textit{non-custodial} pooled stakings (cf. \underline{Tab.\ref{tab:stakingpool}}, $\mathsf{\#Dune}$), and \textit{custodial} stakings (\underline{Tab.\ref{tab:staking-custodial}}). We further provide their comparisons in \underline{Tab.\ref{tab:staking-comparison}} and offer the mappings between non-custodial staking providers and current PoS blockchains (\underline{Tab.\ref{tab:staking-blockchain}}, as demonstrated in \underline{Apx.\ref{apx-stake}}). 

\begin{table}[!hbt]
\caption{Staking services (\textit{non-custodial}, Pooled)}\label{tab:stakingpool}
\renewcommand\arraystretch{1.1}
\begin{center}
\resizebox{1\linewidth}{!}{
\begin{tabular}{lc|ccc|cccccccc|ccc} 

         \multicolumn{1}{c}{\textbf{Pools}} & 
         \multicolumn{1}{c}{\textbf{Mini.}} & 
         \rotatebox{90}{\cellcolor{gray!10}\textbf{Browser}} & 
         \rotatebox{90}{\cellcolor{gray!10}\textbf{Wallet}}  &  
         \rotatebox{90}{\cellcolor{gray!10}\textbf{GUI}} &   
         \rotatebox{90}{\cellcolor{gray!10}\textbf{\makecell{Open-source}}}  &  
         \rotatebox{90}{\cellcolor{gray!10}\textbf{Audited}} & \rotatebox{90}{\cellcolor{gray!10}\textbf{\makecell{Bug bounty}}} & 
         \rotatebox{90}{\cellcolor{gray!10}\textbf{\cellcolor{gray!10}\makecell{Battle-tested}}} &
         \rotatebox{90}{\cellcolor{gray!10}\textbf{Trustless}}& 
         \rotatebox{90}{\cellcolor{gray!10}\textbf{Permissionless}}& 
         \rotatebox{90}{\cellcolor{gray!10}\textbf{\makecell{Cons. diversity}}} & 
         \rotatebox{90}{\cellcolor{gray!10}\textbf{\makecell{Liqui. token}}} &
         \rotatebox{90}{\cellcolor{gray!10}\textbf{\makecell{Validators}}} &  
         \rotatebox{90}{\cellcolor{gray!10}\textbf{\makecell{Mark. share}}} & 
         \rotatebox{90}{\cellcolor{gray!10}\textbf{\makecell{Comissions}}} 
         \\
        \cmidrule{1-8}\cmidrule{13-14}

        \cellcolor{gray!10} \hlhref{https://lido.fi/}{Lido} & \cellcolor{gray!10} Any & \cellcolor{gray!10} \cmark & \cellcolor{gray!10} \cmark  & \cellcolor{gray!10} \cmark & \cellcolor{gray!10} \cmark & \cellcolor{gray!10} \cmark & \cellcolor{gray!10} \cmark & \cellcolor{gray!10} \cmark & \cellcolor{gray!10} \cmark & \cellcolor{gray!10} - &  \cellcolor{gray!10} \cmark & \cellcolor{gray!10} \cmark  & \cellcolor{gray!10} 30.7k & \cellcolor{gray!10} 32\%   & \cellcolor{gray!10} 10\% \\ 
        
        \cellcolor{gray!10} \hlhref{https://rocketpool.net/}{Rocket pool}  & \cellcolor{gray!10} 0.01 & \cellcolor{gray!10} \cmark & \cellcolor{gray!10} & \cellcolor{gray!10} \cmark &\cellcolor{gray!10} \cmark &\cellcolor{gray!10} \cmark &\cellcolor{gray!10} \cmark & \cellcolor{gray!10} \cmark &\cellcolor{gray!10} \cmark &\cellcolor{gray!10} \cmark & \cellcolor{gray!10} \cmark & \cellcolor{gray!10} \cmark & \cellcolor{gray!10} 25k & \cellcolor{gray!10} 3.1\% & \cellcolor{gray!10} 14\%  \\ 

        \cellcolor{gray!10} \hlhref{https://stake.fish/}{Stakefish} & \cellcolor{gray!10} 0.1 &  \cellcolor{gray!10} \cmark &  \cellcolor{gray!10} \cmark  &  \cellcolor{gray!10} \cmark &  \cellcolor{gray!10} - &  \cellcolor{gray!10} \cmark &  \cellcolor{gray!10} - &  \cellcolor{gray!10} \cmark &  \cellcolor{gray!10} \cmark &  \cellcolor{gray!10} - &   \cellcolor{gray!10} \cmark &  \cellcolor{gray!10} - &  \cellcolor{gray!10} 23k &  \cellcolor{gray!10} 2.7\% &  \cellcolor{gray!10} 0.1ETH \\ 

        \cellcolor{gray!10} \hlhref{https://staked.us/}{Staked} & \cellcolor{gray!10} Any & \cellcolor{gray!10} \cmark & \cellcolor{gray!10} - &\cellcolor{gray!10} \cmark  & \cellcolor{gray!10} -  & \cellcolor{gray!10} - & \cellcolor{gray!10} - & \cellcolor{gray!10} - & \cellcolor{gray!10} \cmark & \cellcolor{gray!10} \cmark & \cellcolor{gray!10}  \cmark & \cellcolor{gray!10} - &  \cellcolor{gray!10} 21k &  \cellcolor{gray!10} 2.5\% &  \cellcolor{gray!10} - \\ 

        \cellcolor{gray!10} \hlhref{https://p2p.org/}{P2P} & \cellcolor{gray!10} Any & \cellcolor{gray!10} \cmark & \cellcolor{gray!10} - &\cellcolor{gray!10} \cmark  & \cellcolor{gray!10} -  & \cellcolor{gray!10} - & \cellcolor{gray!10} - & \cellcolor{gray!10} - & \cellcolor{gray!10} \cmark & \cellcolor{gray!10} \cmark & \cellcolor{gray!10}  \cmark & \cellcolor{gray!10} -   & \cellcolor{gray!10} 8k & \cellcolor{gray!10} 0.9\% & \cellcolor{gray!10} 10\%  \\

        \cellcolor{gray!10} \hlhref{https://stakewise.io/}{StakeWise} & \cellcolor{gray!10} Any & \cellcolor{gray!10} \cmark & \cellcolor{gray!10} - & \cellcolor{gray!10} \cmark & \cellcolor{gray!10} \cmark & \cellcolor{gray!10} \cmark & \cellcolor{gray!10} \cmark & \cellcolor{gray!10} \cmark & \cellcolor{gray!10} \cmark & \cellcolor{gray!10} - & \cellcolor{gray!10} \cmark & \cellcolor{gray!10} \cmark & \cellcolor{gray!10} 6k & \cellcolor{gray!10} 0.7\% & \cellcolor{gray!10} 10\%\\ 

        \cellcolor{gray!10} \hlhref{https://www.stafi.io/}{StaFi} & \cellcolor{gray!10} 0.01 & \cellcolor{gray!10}  \cmark & \cellcolor{gray!10}  - & \cellcolor{gray!10}  \cmark & \cellcolor{gray!10}  \cmark & \cellcolor{gray!10}  \cmark & \cellcolor{gray!10}  - & \cellcolor{gray!10}  \cmark & \cellcolor{gray!10}  - & \cellcolor{gray!10}  \cmark & \cellcolor{gray!10}  - &\cellcolor{gray!10}   \cmark &\cellcolor{gray!10} 5k  & \cellcolor{gray!10} 0.6\%  & \cellcolor{gray!10} 10\% \\

        \cellcolor{gray!10} \hlhref{https://www.ankr.com/staking/stake/}{Ankr} & \cellcolor{gray!10}Any &\cellcolor{gray!10} \cmark &\cellcolor{gray!10} \cmark  &\cellcolor{gray!10} \cmark &\cellcolor{gray!10} \cmark &\cellcolor{gray!10} \cmark &\cellcolor{gray!10} \cmark & \cellcolor{gray!10}\cmark &\cellcolor{gray!10} - &\cellcolor{gray!10} - & \cellcolor{gray!10} - &\cellcolor{gray!10} \cmark &\cellcolor{gray!10} 1.7k &\cellcolor{gray!10} 0.2\% & \cellcolor{gray!10} 10\% \\

        \cellcolor{gray!10} \hlhref{https://www.rockx.com/staking/ethereum}{RockX} & \cellcolor{gray!10} Any &\cellcolor{gray!10}  \cmark &\cellcolor{gray!10}  - & \cellcolor{gray!10} \cmark &\cellcolor{gray!10}  \cmark &\cellcolor{gray!10}  \cmark & \cellcolor{gray!10} - & \cellcolor{gray!10} \cmark & \cellcolor{gray!10} \cmark & \cellcolor{gray!10} - &\cellcolor{gray!10}  - & \cellcolor{gray!10} \cmark &\cellcolor{gray!10} 0.13k &\cellcolor{gray!10}  - &\cellcolor{gray!10}  20\%  \\

        \cellcolor{gray!10} \hlhref{https://stakingfacilities.com/}{StakingFacilities} 
        & \cellcolor{gray!10} Any & \cellcolor{gray!10} \cmark & \cellcolor{gray!10} - &\cellcolor{gray!10} \cmark  & \cellcolor{gray!10} -  & \cellcolor{gray!10} - & \cellcolor{gray!10} - & \cellcolor{gray!10} - & \cellcolor{gray!10} \cmark & \cellcolor{gray!10} \cmark & \cellcolor{gray!10}  \cmark & \cellcolor{gray!10} - & \cellcolor{gray!10} - &  \cellcolor{gray!10} -  &   \cellcolor{gray!10} 10\% \\

         \cellcolor{gray!10} \hlhref{https://www.stakewith.us/}{StakeWithUs}& 
        \cellcolor{gray!10} Any & \cellcolor{gray!10} \cmark & \cellcolor{gray!10} -  & \cellcolor{gray!10} - & \cellcolor{gray!10} \cmark & \cellcolor{gray!10} \cmark & \cellcolor{gray!10} \cmark & \cellcolor{gray!10} \cmark & \cellcolor{gray!10} \cmark & \cellcolor{gray!10} - &  \cellcolor{gray!10} \cmark & \cellcolor{gray!10} \cmark & 
        \cellcolor{gray!10} - &  \cellcolor{gray!10} - &  \cellcolor{gray!10}  10\% \\ 
        
        \cellcolor{gray!10} \hlhref{https://stakin.com/}{Stakin} & \cellcolor{gray!10} Any & \cellcolor{gray!10} \cmark & \cellcolor{gray!10} - &\cellcolor{gray!10} \cmark  & \cellcolor{gray!10} -  & \cellcolor{gray!10} - & \cellcolor{gray!10} - & \cellcolor{gray!10} - & \cellcolor{gray!10} \cmark & \cellcolor{gray!10} \cmark & \cellcolor{gray!10}  \cmark & \cellcolor{gray!10} -  &\cellcolor{gray!10} -  &\cellcolor{gray!10} - & \cellcolor{gray!10} 10\%\\

        \cmidrule{5-7}
         \multicolumn{1}{c}{} & \multicolumn{1}{c}{} & \multicolumn{3}{c}{\cellcolor{gray!10}\textbf{Plugin} }   &   \multicolumn{8}{c}{\cellcolor{gray!10}\rotatebox{0}{\textbf{Features}}} &   \multicolumn{3}{c}{\cellcolor{gray!10}\rotatebox{0}{\textbf{Etereum 2.0}}}  \\

\end{tabular}
}
\end{center}
\end{table}

\section{Warm-up Construction}
\label{sec-construction}

\subsection{Delegation Modeling}

We will focus on the delegation phase before validators enter the consensus procedures.
Consider that in a finite set of participants $P$, each participant chooses to be either a {\it user} or a {\it validator}.
$P$ is therefore divided into two subsets: one consists of users $A=\{a_1,\dots, a_n\}$ ($|A|=n$), and the other of validators $V=\{v_1, \dots, v_m\}$ ($|V|=m$).
Hence, we have $A\cup V=P$.
Then, each {\it user} would like to choose a {\it validator} among $V$ to delegate to.
If user $a_i$ chooses to delegate to validator $v_j$, $a_i$ would also decide a weight $t_{i}$ attached to their delegation, which is called the number of {\it tokens} that $a_i$ delegates to $v_j$.
A {\it token profile} $\mathbf{t}=(t_1,\dots, t_n)$ records all users' tokens that they delegate, and $\mathcal{T}$ is the set of all token profiles.
We call $a_i$'s delegation and the token attached to their delegation $a_i$'s {\it delegation strategy}.
Note that we assume that each user cannot delegate to multiple validators.
Note also that for user $a_i$ and validator $v_j$, $t_{ij}=0$ indicates $d_i=0$.

We call $a_i$'s choice of validator the {\it delegation} of $a_i$, denoted as $d_i\in V\cup\{0\}$.
$d_i=v_j$ if $a_i$ delegates to validator $v_j$, and if $d_i=0$, $a_i$ does not delegate to any validator.
Then, a {\it delegation profile} $\mathbf{d}=(d_1,\dots, d_n)$ records each user's delegation.
Let $\mathcal{D}$ be all delegation profiles.
Given a delegation profile $\mathbf{d}$, for each validator $v_j\in V$, let $d(v_j)=\{a_i\in A\mid d_i=v_j\}$ denote the set of users who delegate to $v_j$.
We call $(\mathbf{d},\mathbf{t})$ a {\it strategy profile}.

Users decide their delegation strategies based on two factors: (i) the commission cost by delegating to a validator, and (ii) whether a validator would stay in the market during the slot to validate the next block.
Suppose that user $a_i$ delegates $t_{ij}$ tokens to validator $v_j$ whose commission is $c_j\in [0,1]$, then, $a_i$ would spend $t_{ij}(1+c_j)$ in total to accomplish the delegation.

For each validator $v_j\in V$, we have {\it a priori} probability of events $\theta_j=1$ and $\theta_j=0$, denoting that $v_j$ would stay in the market during the slot, and not, respectively.
The probability of event $\theta_j=1$ is called the {\it integrity} of $v_j$, denoted as $p_j$, and we have that $\Pr(\theta_j=0)= 1-\Pr(\theta_j=1)=1-p_j$.
However, this priori information is not observable by the public, and instead, it is revealed by public {\it evidence} $e_j\in \{0,1\}$ with noise, e.g., the reputation of validators' brands.
$e_j=1$ indicates that the evidence signal shows $v_j$ would stay in the market, while $e_j=0$ indicates that the signal shows $v_j$ would leave.
Then, we use $z_j=\Pr(e_j=1\mid \theta_j=1)=\Pr(e_j=0\mid \theta_j=0)$, i.e., the probability that $v_j$'s evidence truthfully reveals a priori integrity of $v_j$, to denote the {\it quality} of the noisy evidence $e_j$.
In other words, $z_j$ implies how accurate $e_j$ can reveal $\Pr(\theta_j)$.



\subsection{A Probabilistic Model of Trust}
\label{subsec-trust}

Consider that evidence $e_j$ of each validator $v_j$ is observable to the public.
However, users can only act according to the evidence with noise.
Let $q_{ij}=\Pr(\theta_{ij}\mid e_j=1)$ be the {\it accuracy} of user $a_i$'s choice of validator $v_j$ based on evidence $e_j$.
This definition denotes the probability of event $\theta_{ij}: d_i=v_j$
conditioned on $e_j=1$.
On the other hand, let $\bar{q}_{ij}=\Pr(\theta_{ij}\mid e_j=0)$ denote the {\it error} of $a_i$'s choice of $v_j$ based on evidence $e_j$.
Note that we assume that users' accuracies and errors are independent, i.e., for each pair of users $a_i, a_{i'}\in A$ and each validator $v_j\in V$, $\Pr(\theta_{ij},\theta_{i'j}\mid e_j)=\Pr(\theta_{ij}\mid e_j)\Pr(\theta_{i'j}\mid e_j)$.

\begin{figure}[t]
    \centering
    \begin{tikzpicture}
    \node[circle,draw] (1) at (2,3) {$\theta_j$};
    \node[circle,draw] (2) at (2,1.5) {$e_j$};
    \node[circle,draw] (3) at (0,0) {$\theta_{1j}$};
    \node[circle,draw] (4) at (1.5,0) {$\theta_{2j}$};
    \node (5) at (3,0) {$\dots$};
    \node[circle,draw] (6) at (4,0) {$\theta_{nj}$};
    \path[->] (1) edge (2);
    \path[->] (2) edge (3);
    \path[->] (2) edge (4);
    \path[->] (2) edge (6);
    \end{tikzpicture}
    \caption{The influence relationship between $\theta_j$, $e_j$ and $\theta_{ij}$ follows this Bayesian network.}
    \label{fig:bayes}
\end{figure}
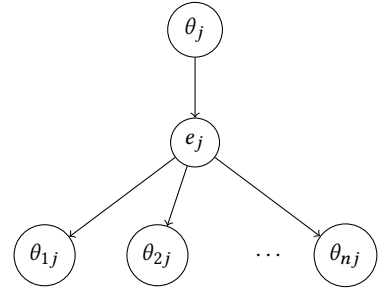

We assume that this noise model satisfies the properties of a Bayesian network shown in \underline{Fig.\ref{fig:bayes}}.
That is, users' accuracies are not conditioned on a priori $\Pr(\theta_j)$.
We remark that such probabilistic structure is also used in the literature of jury theorem, where voters' voting competencies are correlated due to their shared evidence, e.g., \cite{evidence2004}.
Then, for user $a_i$ and validator $v_j$, given $p_j$, $q_j$ and $q_{ij}$, if $a_i$ delegates to $v_j$, $a_i$'s {\it trust} on $v_j$'s integrity is updated based on their behavior $\theta_{ij}$ as:
\begin{align}\label{eq:singlebelief}
&T_{ij}=\Pr(\theta_j=1\mid \theta_{ij})=\frac{\mathsf{B}(a_i,\theta_j=1)}{\mathsf{B}(a_i,\theta_j=1)+\mathsf{B}(a_i,\theta_j=0)}\\
=&\frac{q_{ij}z_jp_j + \bar{q}_{ij}(1-z_j)p_j}{q_{ij}z_jp_j + \bar{q}_{ij}(1-z_j)p_j+q_{ij}(1-z_j)(1-p_j) + \bar{q}_{ij}z_j(1-p_j)},
\end{align}
where
\begin{align*}
\mathsf{B}(a_i,\theta_j=k)= & \Pr(\theta_{ij}\mid e_j=1)\Pr(e_j=1\mid \theta_j=k)\Pr(\theta_j=k)+\\ & \Pr(\theta_{ij}\mid e_j=0)\Pr(e_j=0\mid \theta_j=k)\Pr(\theta_j=k),
\end{align*}
such that $k\in \{0,1\}$, by the Bayesian chain rule.

\textcolor{black}{$\mathsf{Equation~\ref{eq:singlebelief}}$} illustrates a user's trust on a validator based on their own delegation behavior.
Assume that the delegation profile and token profile is observable to the public.
This trust can then be updated based on the observation of the other users' behavior.
Given a delegation profile $\mathbf{d}$, user $a_i$ can update their trust on validator $v_j$ based on the users who delegate to $v_j$.
Formally, given validator $v_j$'s integrity $p_j$, evidence $e_j$'s quality $z_j$ and each user $a_k$'s accuracy $q_{kj}$ and error $\bar{q}_{kj}$ (where $a_k\in A$), the trust of $a_i$ on $v_j$ is a function $T_{ij}:\mathcal{D}\rightarrow [0,1]$, defined as
\begin{align}\label{eq:trust}
T_{ij}(\mathbf{d})= & \Pr(\theta_j=1\mid \wedge_{a_\ell \in d(v_j)}\theta_{\ell j})\\
= & \frac{\mathsf{B}(\{a_\ell\mid d_\ell=v_j\},\theta_j=1)}{\mathsf{B}(\{a_\ell\mid d_\ell=v_j\},\theta_j=1)+\mathsf{B}(\{a_\ell\mid d_\ell=v_j\},\theta_j=0)},
\end{align}
where
\begin{align*}
&\mathsf{B}(\{a_\ell \mid d_\ell=v_j\},\theta_j=1)\\
= & \Pr(\wedge_{a_\ell \in d(v_j)} \theta_{\ell j}\mid e_j=1)\Pr(e_j=1\mid \theta_j=1)\Pr(\theta_j=1)+\\
& \Pr(\wedge_{a_\ell \in d(v_j)} \theta_{\ell j}\mid e_j=0)\Pr(e_j=0\mid \theta_j=1)\Pr(\theta_j=1)\\
= & \prod_{a_\ell \in d(v_j)}\Pr(\theta_{\ell j}\mid e_j=1)\Pr(e_j=1\mid \theta_j=1)\Pr(\theta_j=1)+\\
& \prod_{a_\ell \in d(v_j)}\Pr(\theta_{\ell j}\mid e_j=0)\Pr(e_j=0\mid \theta_j=1)\Pr(\theta_j=1)\\
= & \prod_{a_\ell \in d(v_j)}q_{\ell j}q_jp_j+\prod_{a_\ell \in d(v_j)}\bar{q}_{\ell j}(1-q_j)p_j,
\end{align*}
and similarly
\begin{align*}
&\mathsf{B}(\{a_\ell \mid d_\ell=v_j\},\theta_j=0)\\
= & \Pr(\wedge_{a_\ell \in d(v_j)} \theta_{\ell j}\mid e_j=1)\Pr(e_j=1\mid \theta_j=0)\Pr(\theta_j=0)+\\
& \Pr(\wedge_{a_\ell \in d(v_j)} \theta_{\ell j}\mid e_j=0)\Pr(e_j=0\mid \theta_j=0)\Pr(\theta_j=0)\\
= & \prod_{a_\ell \in d(v_j)}q_{\ell j}(1-z_j)(1-p_j)+\prod_{a_\ell \in d(v_j)}\bar{q}_{\ell j}z_j(1-p_j).
\end{align*}
Observe that all users who delegate to the same validator have the same trust in the validator.

In \underline{Sec.\ref{sec-game}}, for the feasibility of theoretical analysis, we will be working on a simplified class of the above setting: (1) for each validator $v_j$, the evidence has perfect quality, i.e., $z_j=\Pr(e_j=1\mid \theta_j=1)=\Pr(e_j=0\mid \theta_j=0)=1$;
(2) each user $a_i$ has the same accuracy/error on each validator $v_j$, i.e., $q=q_{ij}$ and $\bar{q}=\bar{q}_{ij}$ for all $a_i\in A$ and $v_j\in V$.
Then, given a profile $\mathbf{d}$, we simplify the components of \textcolor{black}{$\mathsf{Equation~\ref{eq:trust}}$} as: $\mathbf{B}(\{a_\ell \mid d_\ell=v_j\},\theta_j=1) = q^kp_j$ and $\mathbf{B}(\{a_\ell \mid d_\ell=v_j\},\theta_j=0) = \bar{q}^k(1-p_j)$, where $k=|d(v_j)|$, and therefore, \underline{Eq.\ref{eq:trust}} becomes
\begin{align}\label{eq:simpletrust}
T_{ij}(\mathbf{d})=\frac{q^kp_j}{q^kp_j + \bar{q}^k(1-p_j)}.
\end{align}
We denote this subclass the {\it Homogeneous User and Perfect Evidence} (HUPE) trust.

\smallskip
\noindent\textbf{Observations in the HUPE trust class.}
We first show that, based on this probabilistic model, when a user cannot access the others' delegation strategies, this user tends to delegate to a validator with higher integrity, if they are more likely to make correct decisions.

\begin{lem}\label{le:tobettervalid}
For each pair of validators $v_j, v_{j'}\in V$ ($j\not= j'$) with $p_j > p_{j'}$ and each user $a_i\in A$, $\Pr(\theta_{ij})>\Pr(\theta_{ij'})$ if $q > \bar{q}$, but $\Pr(\theta_{ij})>\Pr(\theta_{ij'})$ if $q < \bar{q}$.
\end{lem}

\begin{prf}
For $a_i$ and $v_j$, we have that
\begin{align*}
\Pr(\theta_{ij}) & =\Pr(\theta_{ij}\mid e_j=1)\Pr(\theta_j=1)+\Pr(\theta_{ij'}\mid e_j=0)\Pr(\theta_j=0)\\
& = qp_j+\bar{q}(1-p_j)\\
& = \bar{q}+p_j(q-\bar{q}).
\end{align*}
Similarly, we have that $\Pr(\theta_{ij'})=\bar{q}+p_{j'}(q_-\bar{q})$.

\noindent Then, if $q>\bar{q}$, we have that $\Pr(\theta_{ij})>\Pr(\theta_{ij'})$ since $p_j>p_{j'}$.
On the other hand, if $q<\bar{q}$, we have $\Pr(\theta_{ij})<\Pr(\theta_{ij'})$. \qed
\end{prf}
In practice, this may be because users can be correctly directed by the evidence, e.g., the reputation of validators.

\smallskip
Next, we demonstrate that when users possess a higher level of accuracy compared to errors, for a specific user group size, they exhibit more trust in validators with higher integrity.

\begin{lem}\label{le:tobettervali}
Given a pair of delegation profiles $\mathbf{d}$ and $\mathbf{d}'$ and a pair of validators $v_j$ and $v_{j'}$, such that $d(v_j)=d'(d_{j'})$ and $p_{j'}>p_j$, we have that for all $a_i\in d(v_j)$, $T{ij'}(\mathbf{d}')>T_{ij}(\mathbf{d})$ if $q>\bar{q}$.
\end{lem}

\begin{prf}
Let $|d(v_j)|=|d'(v_{j'})|=k$.
We write the trusts of $a_i$ on $v_j$ and $v_{j'}$ in delegation profiles $\mathbf{d}$ and $\mathbf{d}'$ as:
\begin{align}\label{eq:le2:tij}
T_{ij}(\mathbf{d})=\frac{p_jq^k}{p_jq^k+(1-p_j)\bar{q}^k}=\frac{p_jq^k}{p_j(q_k-\bar{q}^k)+\bar{q}^k},
\end{align}
and
\begin{align}\label{eq:le2:tij'}
T_{ij'}(\mathbf{d}')=\frac{p_{j'}q^k}{p_{j'}(q_k-\bar{q}^k)+\bar{q}^k}.
\end{align}

Dividing \underline{Eq.\ref{eq:le2:tij'}} by \underline{Eq.\ref{eq:le2:tij}} and we have that $$T_{ij'}(\mathbf{d}')/T_{ij}(\mathbf{d})=\frac{p_jp_{j'}(q^k-\bar{q}^k)+p_{j'}\bar{q}^k}{p_jp_{j'}(q^k-\bar{q}^k)+p_{j}\bar{q}^k}.$$
Since $p_{j'}>p_j$, and both $p_jp_{j'}(q^k-\bar{q}^k)+p_{j'}\bar{q}^k$ and $p_jp_{j'}(q^k-\bar{q}^k)+p_{j}\bar{q}^k$ are positive, we have that $T_{ij'}(\mathbf{d}')/T_{ij}(\mathbf{d})>1$ which implies that $T_{ij'}(\mathbf{d}')>T_{ij}(\mathbf{d})$. \qed
\end{prf}

Our last observation on trust shows that more users delegating to a validator enhances the trust of users on the validator if users have higher accuracy than error, however, it undermines the trust if users have lower accuracy than error.

\begin{lem}\label{le:monotone}
Given a pair of delegation profiles $\mathbf{d}$ and $\mathbf{d}'$, such that for validator $v_j$, $d'(v_j)=d(v_j)\cup \{a_k\}$ ($a_k\notin d(v_j)$), we have that for each $a_i\in d(v_j)$, $T_{ij}(\mathbf{d}')>T_{ij}(\mathbf{d})$ if $q > \bar{q}$, $T_{ij}(\mathbf{d}')<T_{ij}(\mathbf{d})$ if $q < \bar{q}$, and $T_{ij}(\mathbf{d}')=T_{ij}(\mathbf{d})$ if $q=\bar{q}$.
\end{lem}

\begin{prf}
We prove this lemma by showing that if $q > \bar{q}$, $T_{ij}(\mathbf{d})/T_{ij}(\mathbf{d}')<1$, if $q < \bar{q}$, $T_{ij}(\mathbf{d})/T_{ij}(\mathbf{d}')>1$, and if $q = \bar{q}$, $T_{ij}(\mathbf{d})/T_{ij}(\mathbf{d}')=1$.

First, we notice that since each user's accuracy and error are identical for each validator, $T_{ij}(\cdot)$ solely depends on $p_j$ and the number of users delegating to $v_j$.
Let $|d(j)|=k$.
We then have that
\begin{align}\label{eq:td}
T_{ij}(\mathbf{d})=\frac{p_jq^k}{p_jq^k+(1-p_j)\bar{q}^k},
\end{align}
and
\begin{align}\label{eq:td'}
T_{ij}(\mathbf{d}')=\frac{p_jq^{k+1}}{p_jq^{k+1}+(1-p_j)\bar{q}^{k+1}}.
\end{align}
By dividing \underline{Eq.\ref{eq:td}} by \underline{Eq.\ref{eq:td'}}, we have that
\begin{align}\label{eq:td/td'}
T_{ij}(\mathbf{d})/T_{ij}(\mathbf{d}') = \frac{(1-p_j)\bar{q}^{k+1}+p_jq^{k+1}}{(1-p_j)\bar{q}^kq+p_jq^{k+1}}.
\end{align}
Hence, if $q>\bar{q}$, $T_{ij}(\mathbf{d})/T_{ij}(\mathbf{d}')< 1$, and therefore, $T_{ij}(\mathbf{d}')>T_{ij}(\mathbf{d})$.
Reversely, if $q<\bar{q}$, $T_{ij}(\mathbf{d}')<T_{ij}(\mathbf{d})$.
Specially, if $q=\bar{q}$, $T_{ij}(\mathbf{d'})/T_{ij}(\mathbf{d})=1$, which indicates $T_{ij}(\mathbf{d'})=T_{ij}(\mathbf{d})$. \qed
\end{prf}

In other words, given the fact that each user is more likely to make a decision coincident with the observed evidence, i.e., $q > \bar{q}$, users better trust a validator if more users delegate to the validator, but users less trust the validator if fewer users delegate to the validator. A special case is that the trust on a validator is insensitive to the number of delegations if $q=\bar{q}$.

\section{A Game Theoretic Model for Validator Selection}
\label{sec-game}

In this section, we define a game theoretical model in which each user chooses their delegation strategy in order to maximize their expected profit through the block validation slot.

In the rest of this section, we first provide the definition of the game in the general setting introduced in \underline{Sec.\ref{sec-construction}}, and then, we show theoretical analysis in subclasses under the HUPE trust class, e.g., the homogeneous belief class.

\subsection{Validator Selection Game (VSG)}
\label{subsec-gamedef}

\begin{defi}[VSG]\label{defi-vsgame}
    A {\it validator selection game} (VSG) is denoted as a tuple 
$$\g=\langle A, V, \mathbf{q}, \mathbf{p},\mathbf{z}, \mathbf{c}, \mathbf{b}, \mathbf{\Sigma}, r,\mathbf{u} \rangle,$$
where $A$ is a finite set of users and $V$ a finite set of {\it validators}.
$\mathbf{q}=((q_{11},\bar{q}_{11}),\dots, (q_{nm},\bar{q}_{nm}))\in \mathbb{R}^{n\times m\times 2}_{\ge 0}$ is a profile of users accuracies and errors on $m$ validators.
$\mathbf{p}=(p_1,\dots, p_m)\in \mathbb{R}^m_{\ge 0}$ is a {\it integrity profile}.
$\mathbf{z}=(z_1,\dots, z_m)\in \mathbb{R}^m_{\ge 0}$ is a profile of evidence quality. 
$\mathbf{c}=(c_1, \dots, c_m)\in \mathbb{R}^m_{\ge 0}$ is a {\it commission profile}.
$\mathbf{b}=(b_1,\dots, b_n)\in \mathbb{R}^n_{>0}$ is a {\it budget profile}, i.e., the number of tokens that each user can use.
$\Sigma_i\in V\times \mathbb{R}_{\ge 0}$ is the {\it strategy space} of user $a_i$, where $(d_i,t_i)\in \Sigma_i$ denotes that user $a_i$ delegates $t_i$ tokens to validator $v_j$ under the budget constraint $t_i(1+c_j)\le b_i$.
User $a_i$ abstains if $t_i=0$, that is, they do not delegate.
Finally, given the accuracy and error profile and the {\it profit} $r$ (that the system returns when the next block is validated), $u_i: \mathcal{D}\times \mathcal{T}\rightarrow \mathbb{R}$ is the {\it utility function} of user $a_i$. 
\end{defi} 

We define the utility function as follows.

\begin{defi}[Utility] \label{defi-utility}
Given a strategy profile $(\mathbf{d},\mathbf{t})$ and a profit $r$, for each user $a_i$, their utility function is:
\begin{align}\label{eq:utility}
    u_i(\mathbf{d},\mathbf{t})=\frac{rT_{id_i}(\mathbf{d})t_i}{\sum_{j=1}^nT_{jd_j}(\mathbf{d})t_j}-c_{d_i}t_i-(1-T_{id_i}(\mathbf{d}))t_i,
\end{align}
where
$T$ is the trust function described in \underline{\textrm{Sec.\ref{subsec-trust}}}.
\end{defi}

Intuitively, user $a_i$ expects that they delegate to a validator a number of tokens that is the multiplication of the actual delegated token number $t_i$ and the trust $T_{id_i}$, i.e., the probability that they believe the validator would not leave the market.
Then, their utility is a proportion of the total profit $r$ subtracting the commission and the expected loss of tokens (i.e., $(1-T_{id_i}(\mathbf{d}))t_i$).
The proportion is decided by the ratio of the user's expected delegating token number out of the entire expected token number in the pool.

Note that in this section, we assume that, users' accuracy is higher than their error, i.e., $q>\bar{q}$.

\begin{rmk}
In the above definition, we assume that each user's token strategy space lies in the non-negative real space, i.e., $\Sigma_i\in V\times \mathbb{R}_{\ge 0}$ for all $a_i\in A$.
That is, depending on the specific VSG setting, a user's token strategy might be any real number between $0$ and their budget $b_i$.
However, in practice, users' token strategy spaces are usually discrete, e.g., $\Sigma_i=V\times [b_i]$, where $[b_i]=\{0,1,\dots, b_i\}$.
In this work, we consider continuous token strategy spaces mainly because of their simplicity for theoretical analysis.
In \underline{\textrm{Sec.\ref{sec-experiment}}}, we consider the more practical setting of discrete token strategy spaces.
\end{rmk}

To study users' behavior in VSGs, we will be considering the existence and the structure of the well-known game solution, the {\it Nash equilibrium} (NE).

\begin{defi}[Nash equilibrium]\label{defi-nash}
Given a VSG $\g$, a strategy profile $(\mathbf{d},\mathbf{t})$ is a Nash equilibrium if there is no user $a_i\in A$ and their strategy $(d'_i,t'_i)$ such that $(d'_i,t'_i)\not= (d_i,t_i)$, and $u_i((\mathbf{d}_{-i},d'_i),(\mathbf{t}_{-i},t'_i))>u_i(\mathbf{d},\mathbf{t})$, where $\mathbf{d}_{-i}$ and $\mathbf{t}_{-i}$ denotes that the other users than $a_i$ take the delegation strategy and the token strategy in $\mathbf{d}$ and $\mathbf{t}$ respectively.
\end{defi}

We use the following example to illustrate the above definitions.

\begin{exmp}
Consider a VSG with two users $A=\{a_1,a_2\}$ and two validators $V=\{v_1,v_2\}$.
Both users $a_1$ and $a_2$ have the same accuracy and error $(q,\bar{q})=(0.8,0.3)$ on both validators $v_1$ and $v_2$, and we consider both users' budgets are large.
For validators, the commission profile is $\mathbf{c}=(0.2, 0.1)$, the integrity profile is $\mathbf{p}=(0.8,0.6)$.
When the next block is validated, users will receive a large amount of profit $r$.
\medskip

\noindent We first compute both users' trusts on both validators based on all possible delegation profiles by \underline{Eq.\ref{eq:simpletrust}} as follows.
\begin{table}[!hbt]
\caption{Users' trusts on validators. $T_{ij}$ denotes the trust of user $a_i$ on validator $v_j$.}\label{tab:ex:1:trust}
\begin{center}
\begin{tabular}{c|c c} 
 & $d_2=v_1$ & $d_2=v_2$\\
\hline
$d_1=v_1$ & \cellcolor{gray!10}  $T_{11}=0.966, T_{21}=0.966$ & \cellcolor{gray!10} $T_{11}=0.914,T_{22}=0.8$\\
$d_1=v_2$ & \cellcolor{gray!10} $T_{12}=0.8, T_{21}=0.914$ & \cellcolor{gray!10} $T_{12}=0.914,T_{22}=0.914$

\end{tabular}
\end{center}
\end{table}

We then study the utility each user can obtain by taking different strategies.
We first investigate how users decide their number of delegating tokens $\mathbf{t}=(t_1,t_2)$ with their delegation profile fixed as $\mathbf{d}=(d_1,d_2)$.
Let $T_1$ and $T_2$ denote the trust of $a_1$ and $a_2$ on their delegating validators, respectively.
By \underline{Eq.\ref{eq:utility}}, we have that $a_1$ obtains utility of:
\begin{align}\label{eq:ex1:u1}
u_1(\mathbf{d},\mathbf{t})=\frac{rT_1 t_1}{T_1t_1+T_2t_2}-c_{d_1}t_1-(1-T_1)t_1.
\end{align}
Differentiating \underline{Eq.\ref{eq:ex1:u1}} by $t_1$, we have that
\begin{align*}
\frac{du_1}{dt_1}=\frac{rT_2t_2T_1}{(T_1t_1+T_2t_2)^2}-(c_{d_1}+1-T_1).
\end{align*}
Therefore, we have that user $a_1$ optimizes $u_1(\mathbf{d},\mathbf{t})$ by (1) $t^*_1=0$, corresponding to utility of $0$, if $\frac{rT_1}{T_2t_2}-(c_{d_1}+1-T_1)<0$ (i.e., $u_1$ always decreases as $t_1$ increases), or (2) $t^*_1=\sqrt{\frac{rT_2t_2}{w_1T_1}}-\frac{T_2}{T_1}t_2$, by letting $\frac{du_1}{dt_1}=0$, where $w_1=(c_{d_1}+1-T_1)$.
Then, in case (2), we have that the optimal utility $a_1$ can obtain is
\begin{align}\label{eq:ex1:maxu1}
u^*_1(\mathbf{d},(t^*_1,t_2))=r-2\sqrt{\frac{rT_2t_2w_1}{T_1}}+\frac{T_2}{T_1}t_2w_1=\left(\sqrt{\frac{T_2t_2w_1}{T_1}}-\sqrt{r}\right)^2.
\end{align}

Assume that both users enter the market by taking delegation profile $(d_1=v_2, d_2=v_2)$.
Their trusts on $v_2$ are identical since they delegate to the same validator, and we let the trust be $T$.
Then, having $\frac{du_1}{t_1}=\frac{du_2}{t_2}=0$,
we have that both users delegate $t=\frac{r}{4w}$ tokens, otherwise, they have an incentive to alter their token strategy to achieve a higher utility.

Then, we show that strategy profile $((d_1=v_2,d_2=v_2),(t,t))$ is a NE.
First notice that under this strategy profile, each user obtains utility of $\frac{r}{4}>0$, which indicates that no user has an incentive to abstain.
Then, we show that no user has an incentive to delegate to validator $v_1$.
In particular, we show that $a_1$ has no incentive to deviate to $d'_1=v_1$, and a similar reasoning can be developed to show that $a_2$ has no incentive to deviate from $(d_1=v_2, d_2=v_2)$ to $d'_2=v_1$.

Consider strategy profile $((d_1=v_2, d_2=v_2), (t,t))$.
If $a_2$ fixes their strategy $(d_2,t)$ but $a_1$ changes from $d_1=v_2$ to $d'_1=v_1$, the trusts of $a_1$ on $v_2$ before the change and on $v_1$ after the change are identical ($0.914$) as shown in \underline{Tab.\ref{tab:ex:1:trust}}, but the trust of $a_2$ on $v_1$ changes from $0.914$ to $0.8$.
Considering $a_1$ bears a higher commission rate from delegating to $c_2$ to $c_1$, the only changed component in \textcolor{black}{$\mathsf{Equation~\ref{eq:ex1:maxu1}}$} is $w_1T_2$, from $0.17$ to $0.2288$.
Since $u_1(\mathbf{d},\mathbf{t})>0$ in \textsf{Equation}~\ref{eq:ex1:u1}, otherwise $a_1$ would rather abstain,
we have that
\begin{align*}
\frac{rT_1}{T_2t_2}-w_1>\frac{rT_1}{T_1t_1+T_2t_2}-w_1>0,
\end{align*}
which further indicates
$\frac{T_2t_2w_1}{T_1}<r$.

This indicates that, according to \textsf{Equation~\ref{eq:ex1:maxu1}}, user $a_1$ obtains a lower utility by changing from $d_1=v_2$ to $d'_1=v_1$ even by always taking the optimal token strategy because $w_1T_2$ increases.
A similar argument can be developed for the user $a_2$, and therefore, $((d_1=v_2,d_2=v_2),(t,t))$ is a NE.

Observe that in the above NE, users do not delegate to $v_1$ who has a higher integrity.
Instead, a rational user tries to reach a good balance between high reputation and low commission, such as to optimize their utility.
\qed
\end{exmp}

\noindent Next, we investigate the existence and the feature of NE in VSGs under the HUPE trust class, which we call HUPE VSGs.

\subsection{Equilibria in HUPE VSG}
We show the existence and structure of NE in several subclasses of HUPE VSGs, through which we gain insight into how users delegate and use their tokens in markets under certain conditions.

\subsubsection{Single Validator VSG}\label{subsec-single}
\hfill
\medskip

We first study a subclass of the HUPE VSGs: the {\it single validator} VSGs, where there is only one validator in the market.
Then, each user can choose to delegate to this validator with a number of tokens within their budget, or abstain.
As follows, we show that there always exists an NE in each single validator VSG if a necessary condition on users' budgets holds.

\begin{thm}\label{thm:NEsingle}
In a single validator VSG, i.e., there is only one validator, there always exists a NE if for all user $a_i\in A$, their budget $b_i$ satisfies $b_i\ge \frac{(n-1)r}{n^2(1+c-T)}(1+c)$, where $n=|A|$, $r$ is the profit, $c$ is the commission rate of the validator, and $T=\frac{q^np}{q^np+\bar{q}^n(1-p)}$ ($p$ is the validator's integrity, and $q$ and $\bar{q}$ are users' accuracy and error).
\end{thm}

\begin{prf}[Thm.\ref{thm:NEsingle}]
We prove this theorem by showing that the strategy profile $(\mathbf{d}^*,\mathbf{t}^*)=(d_1=\dots=d_n=v, t_1=\dots=t_n=t=\frac{(n-1)r}{n^2(1+c-T)})$, i.e., each user delegates to the only validator $v$ with $t$ tokens, is an NE.

We first show that in strategy profile $(\mathbf{d}^*,\mathbf{t}^*)$, no user has the incentive to deviate by abstaining.
Since each user delegates to $v$, all users' trusts on $v$ are identical, and equal
\begin{align*}
T=\frac{q^np}{q^np+\bar{q}^n(1-p)}.
\end{align*}
For an arbitrary user $a_i\in A$, their utility by taking the above strategy profile is:
\begin{align*}
u_i(\mathbf{d}^*,\mathbf{t}^*)=&\frac{rTt}{\sum_{a_i\in A}Tt}-wt =\frac{r}{n}-wt=\frac{r}{n^2}\ge 0,
\end{align*}
where $w=1+c-T$.
If $a_i$ abstains, they obtain utility of $0$, which indicates that $a_i$ prefers to take $(\mathbf{d}^*,\mathbf{t}^*)$ rather than to abstain.

Then, we show that no user can obtain a higher utility by unilaterally altering their token strategy.
Assume that user $a_i$ alters his token strategy to $t+x$, such that $t+x\in [0,b_i]$.
Let the altered strategy profile be $(\mathbf{d}^*,\mathbf{t}'=(\mathbf{t}^*_{-i},t_i=t+x))$, where $\mathbf{t}^*_{-i}$ denotes that all users except for $a_i$ take $\mathbf{t}^*$.
The utility of $a_i$ becomes:
\begin{align*}
u_i(\mathbf{d}^*,\mathbf{t}')=\frac{r(t+x)}{nt+x}-w(t+x).
\end{align*}
Differentiating $u_i(\mathbf{d}^*,\mathbf{t}')$ with respect to $x$, we have
\begin{align*}
\frac{du_i(\mathbf{d}^*,\mathbf{t}')}{dx}=\frac{(n-1)rt}{(nt+x)^2}-w.
\end{align*}
Substitute $t$ by $\frac{(n-1)r}{n^2w}$, and we obtain
\begin{align*}
\frac{du_i(\mathbf{d}^*,\mathbf{t}')}{dx}=\left(\frac{(n-1)^2r^2-[(n-1)r+nwx]^2}{n^2w}\right)/\left(\frac{(n-1)r+nwx}{nw}\right)^2.
\end{align*}
Observe that when $x=0$, $\frac{du_i(\mathbf{d}^*,\mathbf{t}')}{dx}=0$, and $\frac{du_i((\mathbf{d}^*,\mathbf{t}'))}{dx}>0$ when $x<0$ and $\frac{du_i(\mathbf{d}^*,\mathbf{t}')}{dx}<0$ when $x>0$.
That is, $a_i$ cannot increase their utility from $u_i(\mathbf{d}^*,\mathbf{t}^*)$ by choosing $x\not= 0$.

\noindent This completes the proof.
\qed
\end{prf}

\underline{Thm.\ref{thm:NEsingle}} illustrates that in a single validator VSG, users reach a NE by delegating to the validator with an identical number of tokens if their budgets admit.
Observe that when users' budgets are high, they do not delegate as many as possible.
Instead, compared to the NE token strategy shown in \underline{Thm.\ref{thm:NEsingle}}, if a user's delegating token number is at a lower level, they can improve utility by increasing the delegating number, and if at a high level, they can improve by decreasing the number.
Then, users' strategies eventually converge to the NE given that all users are rational and selfish.

\subsubsection{Homogeneous Validator VSG}\label{subsec-homo}
\hfill
\medskip

We next consider another subclass of the HUPE VSGs, the {\it homogeneous validator VSG}, where each validator has the same integrity and commission rate. The following theorem shows that an NE can always be guaranteed in a homogeneous validator VSG under certain conditions.

\begin{thm}\label{thm:NEhomov}
In a homogenous validator VSG, there always exists an NE, if for each user $a_i\in A$, their budget $b_i$ satisfies $b_i\ge \frac{(n-1)r}{n^2(1+c-T)}(1+c)\ge 1+c$, where $n=|A|$, $r$ is the profit, $c$ is the commission rate of the validator, and $T=\frac{q^np}{q^np+\bar{q}^n(1-p)}$ ($p$ is the validator's integrity, and $q$ and $\bar{q}$ are users' accuracy and error).
\end{thm}

\begin{prf}[Thm.\ref{thm:NEhomov}]
We prove by showing that the strategy profile $(\mathbf{d},\mathbf{t})$, where for each $a_i\in A$, $d_i=v_j$ and $t_i=t=\frac{(n-1)r}{n^2(1+c-T)}$ is a NE, where $v_j\in V$ and $T=\frac{q^np}{q^np+\bar{q}^n(1-p)}$.
Note that $t\ge 1$ by the condition of the theorem.
For an arbitrary user $a_i\in A$, we reason by three exclusive aspects: \textbf{(i)} $a_i$ cannot obtain a higher utility by abstaining; \textbf{(ii)} $a_i$ cannot obtain a higher utility by only altering their token strategy; and \textbf{(iii)}  $a_i$ cannot obtain a higher utility by delegating to another validator.

\smallskip
\noindent\textbf{\textit{Case-}(i).}
Taking $(\mathbf{d},\mathbf{t})$, $a_i$ obtains utility:
\begin{align*}
u_i(\mathbf{d},\mathbf{t})=\frac{r}{n}-wt=\frac{r}{n^2}>0,
\end{align*}
where $w=1+c_j-T$.
If $a_i$ abstains, they obtain utility of $0$, which is lower than $u_i(\mathbf{d},\mathbf{t})$, which implies that $a_i$ would prefers $(\mathbf{d},\mathbf{t})$ over abstention.

\smallskip
\noindent\textbf{\textit{Case-}(ii).}
By the proof of \underline{Thm.\ref{thm:NEsingle}}, we have that this claim holds.

\smallskip
\noindent\textbf{\textit{Case-}(iii).}
Assume that $a_i$ delegate to validator $v_{j'}$ (this includes the case $v_j = v_{j'}$) with token strategy $t'$, forming profile $(\mathbf{d}',\mathbf{t}')=((\mathbf{d}_{-i},d'_i=v_{j'}),(\mathbf{t}_{-i},t'))$, and we have that $a_i$ obtains utility:
\begin{align*}
u_i(\mathbf{d}',\mathbf{t}')=\frac{rT_{ij'}t'}{(n-1)T^{(n-1)}t+T_{ij'}t'}-w_{j'}t',
\end{align*}
where $T^{(n-1)}$
is the trust of the other users than $a_i$ on $v_j$, $T_{ij'}$
is the trust of $a_i$ on $v_{j'}$, and $w_{j'}=1+c_{j'}-T_{ij'}$.
We have the derivative of $u_i(\mathbf{d}',\mathbf{t}')$ with respect to $t'$ is:
\begin{align*}
\frac{du_i(\mathbf{d}',\mathbf{t}')}{dt'} = \frac{r(n-1)T^{(n-1)}tT_{ij'}}{((n-1)T^{(n-1)}t+T_{ij'}t')^2}-w_{ij'}.
\end{align*}
Making the above derivative $0$, we have that the optimal token strategy for $a_i$ is
\begin{align}\label{eq:optimaltoken}
t'=\sqrt{\frac{r(n-1)T^{(n-1)}t}{w_{ij'}T_{ij'}}}-\frac{(n-1)T^{(n-1)}t}{T_{ij'}}.
\end{align}
Then, we have that the optimal utility $a_i$ can obtain by delegating to $v_{j'}$ is:
\begin{align*}
u^*_i= r-2\sqrt{\frac{r(n-1)T^{(n-1)}w_{ij'}}{T_{ij'}}}+ \frac{w_{ij'}(n-1)T^{(n-1)}}{T_{ij'}}.
\end{align*}
Let $X=\sqrt{\frac{(n-1)T^{(n-1)}w_{ij'}}{T_{ij'}}}$, and we have that $u^*_i=(X-\sqrt{r})^2$.

\noindent Compare the two cases: (1) $v_{j'}=v_j$ and (2) $v_{j'}\not= v_j$.
From (1) to (2), $w_{ij'}$ increases because $T_{ij'}<T$.
We also have that, from (1) to (2), $T^{(n-1)}/T_{ij'}$ increases from $1$ to larger than $1$ because $T^{(n-1)}=T_{ij'}$ if $v_{j'}=v_j$, and $T^{(n-1)}>T_{ij'}$ if $v_{ij'}\not=v_j$ by \underline{Lm.\ref{le:monotone}}.
Therefore, $X$ increases from (1) to (2).

\noindent Since $u_i(\mathbf{d}',\mathbf{t}')>0$, otherwise $a_i$ prefers abstain, we have that
\begin{align*}
\frac{rT_{ij'}}{(n-1)T^{(n-1)}t}-w_{ij'}>\frac{rT_{ij'}}{(n-1)T^{(n-1)}t+T_{ij'}t'}-w_{ij'}>0.
\end{align*}
Since by the theorem's condition, $t\ge 1$, we have that
\begin{align*}
\frac{(n-1)T^{(n-1)}w_{ij'}}{T_{ij'}}<r.
\end{align*}
This indicates that, since $X$ increases from (1) to (2), $u^*_i=(X-\sqrt{r})^2$ decreases.
Thus, $a_i$ does not have an incentive to change from $(\mathbf{d},\mathbf{t})$ to delegating to another validator.

\noindent This completes the proof.
\qed
\end{prf}

\begin{figure*}[!hbt]
\centering
\subcaptionbox
{$\epsilon=0.7$ with $1$ round. \label{fig:e7r1}}{\includegraphics[width=4.3cm]{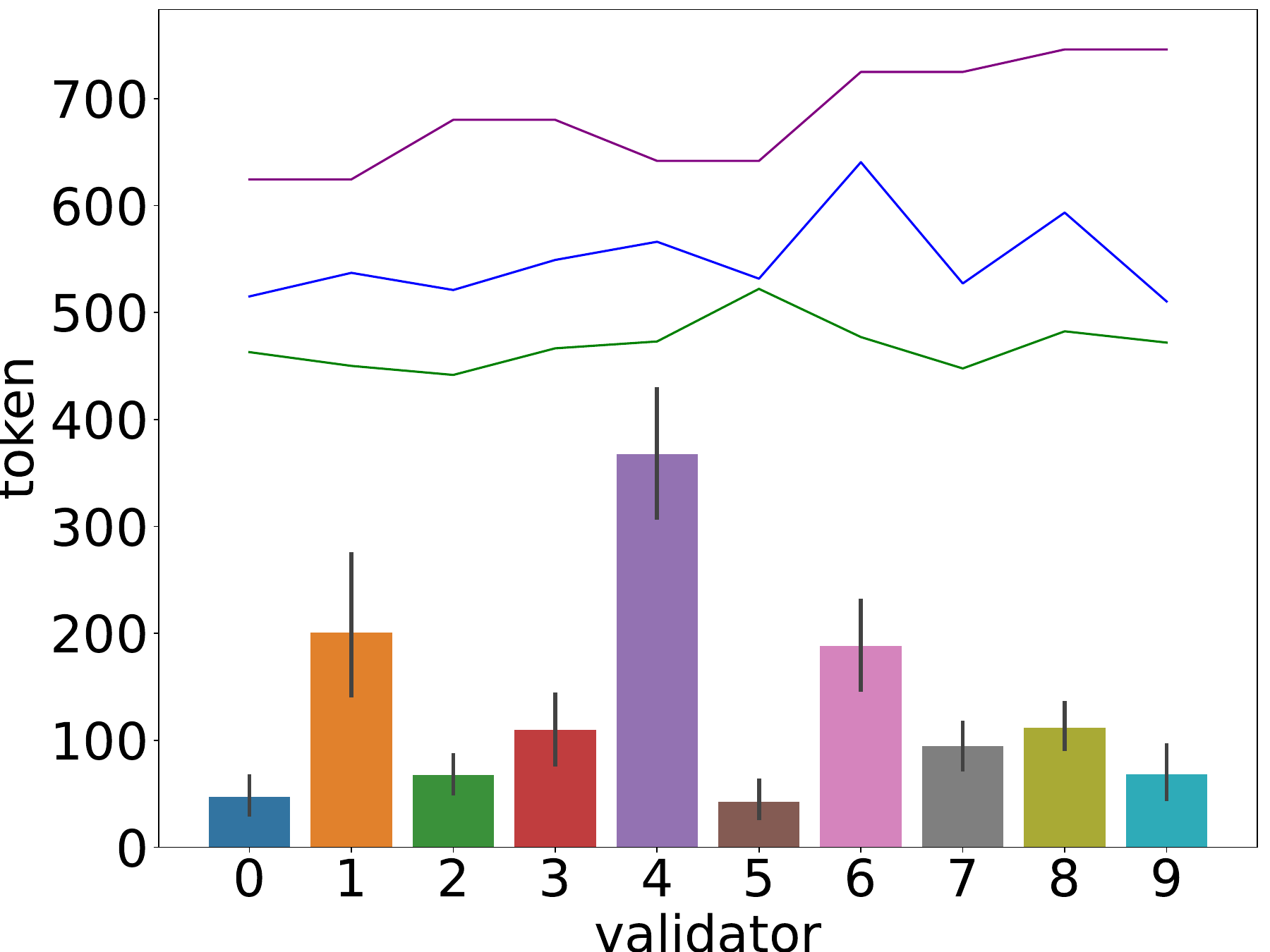}}
\subcaptionbox
{$\epsilon=0.8$ with $1$ round. \label{fig:e8r1}}{\includegraphics[width=4.3cm]{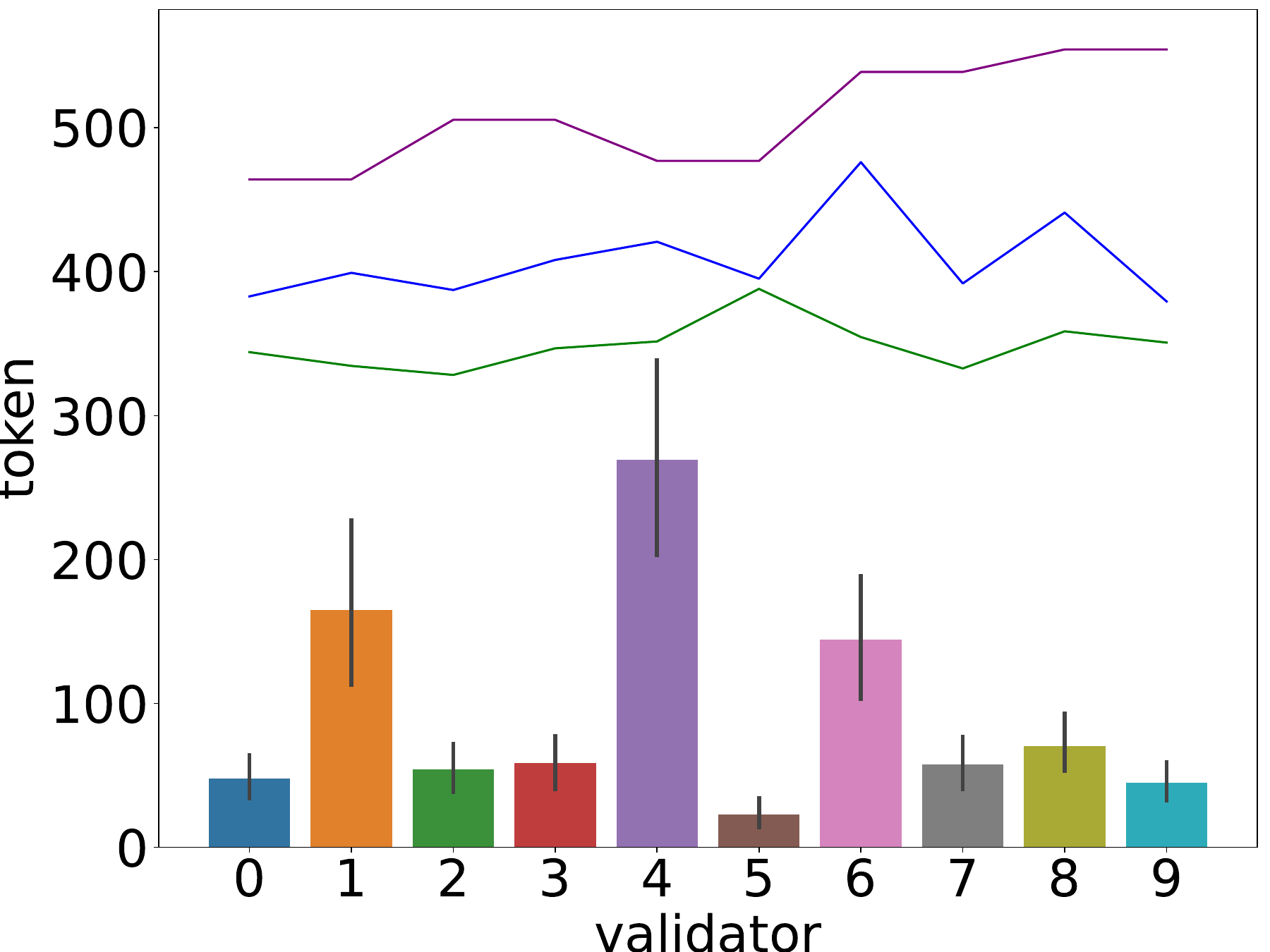}}
\subcaptionbox
{$\epsilon=0.9$ with $1$ round. \label{fig:e9r1}}{\includegraphics[width=4.3cm]{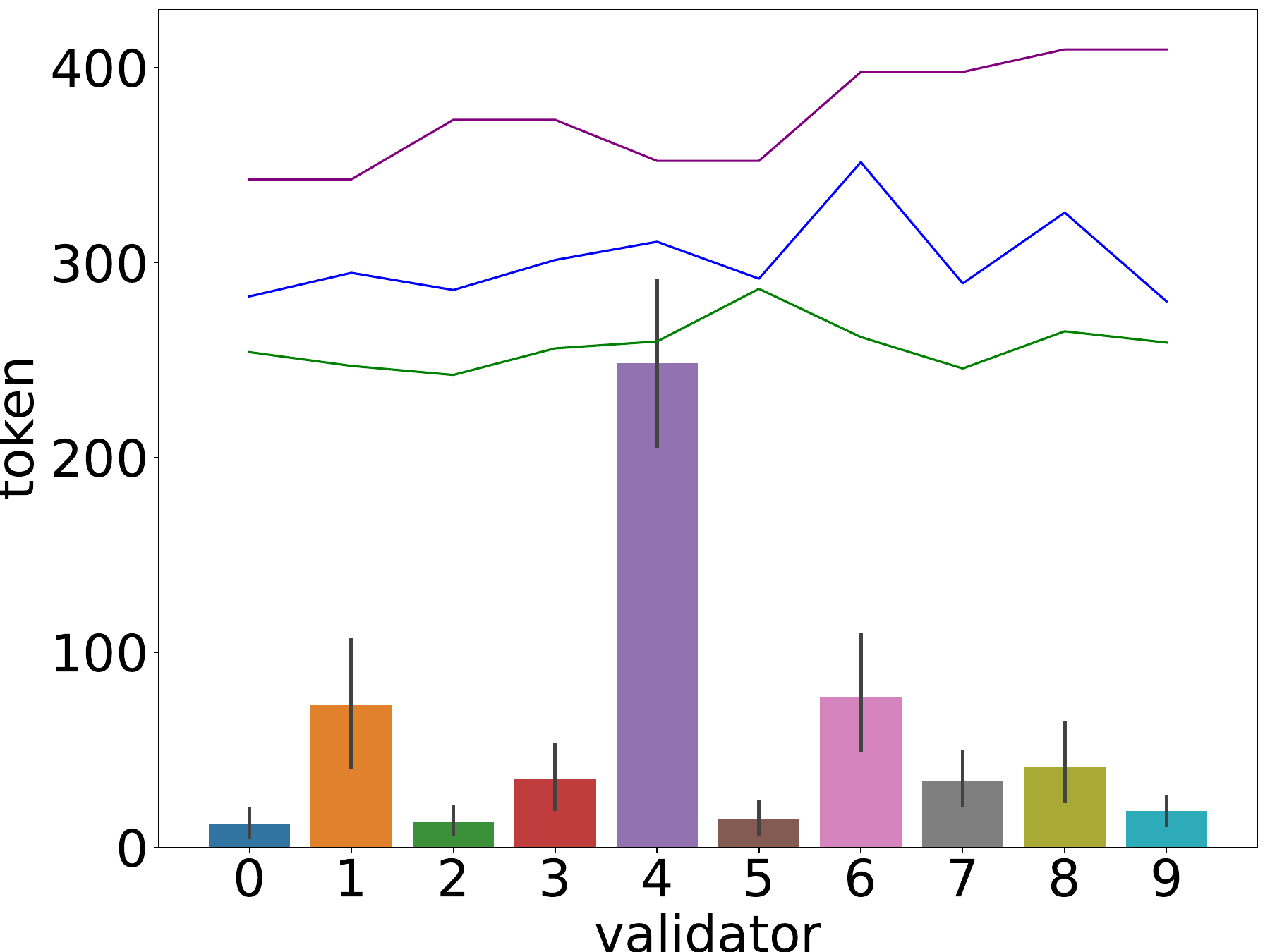}}
\subcaptionbox
{$\epsilon=1$ with $1$ round. \label{fig:e10r1}}{\includegraphics[width=4.3cm]{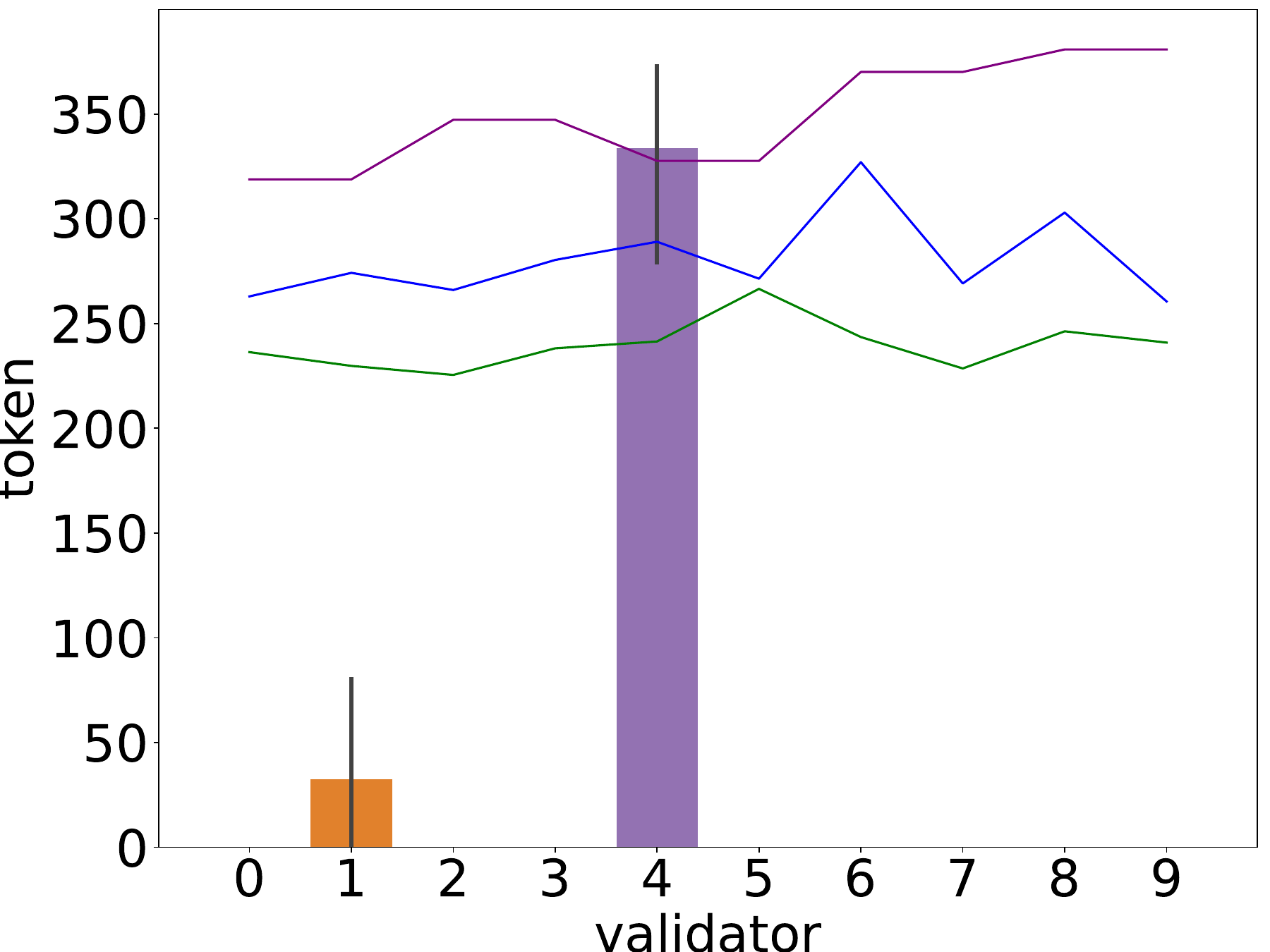}}
\subcaptionbox
{$\epsilon=0.7$ with $2$ round. \label{fig:e7r2}}{\includegraphics[width=4.3cm]{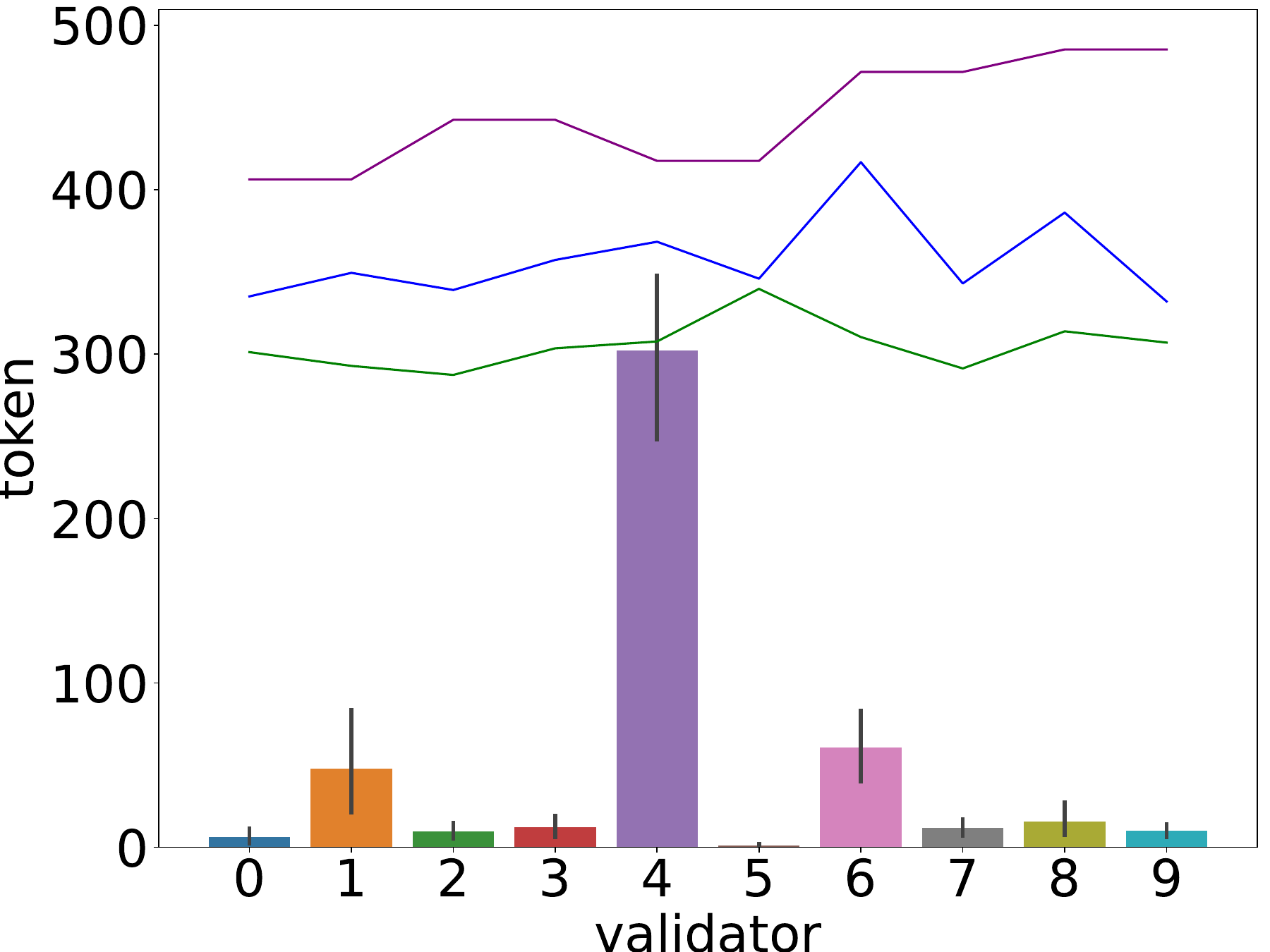}}
\subcaptionbox
{$\epsilon=0.8$ with $2$ round. \label{fig:e8r2}}{\includegraphics[width=4.3cm]{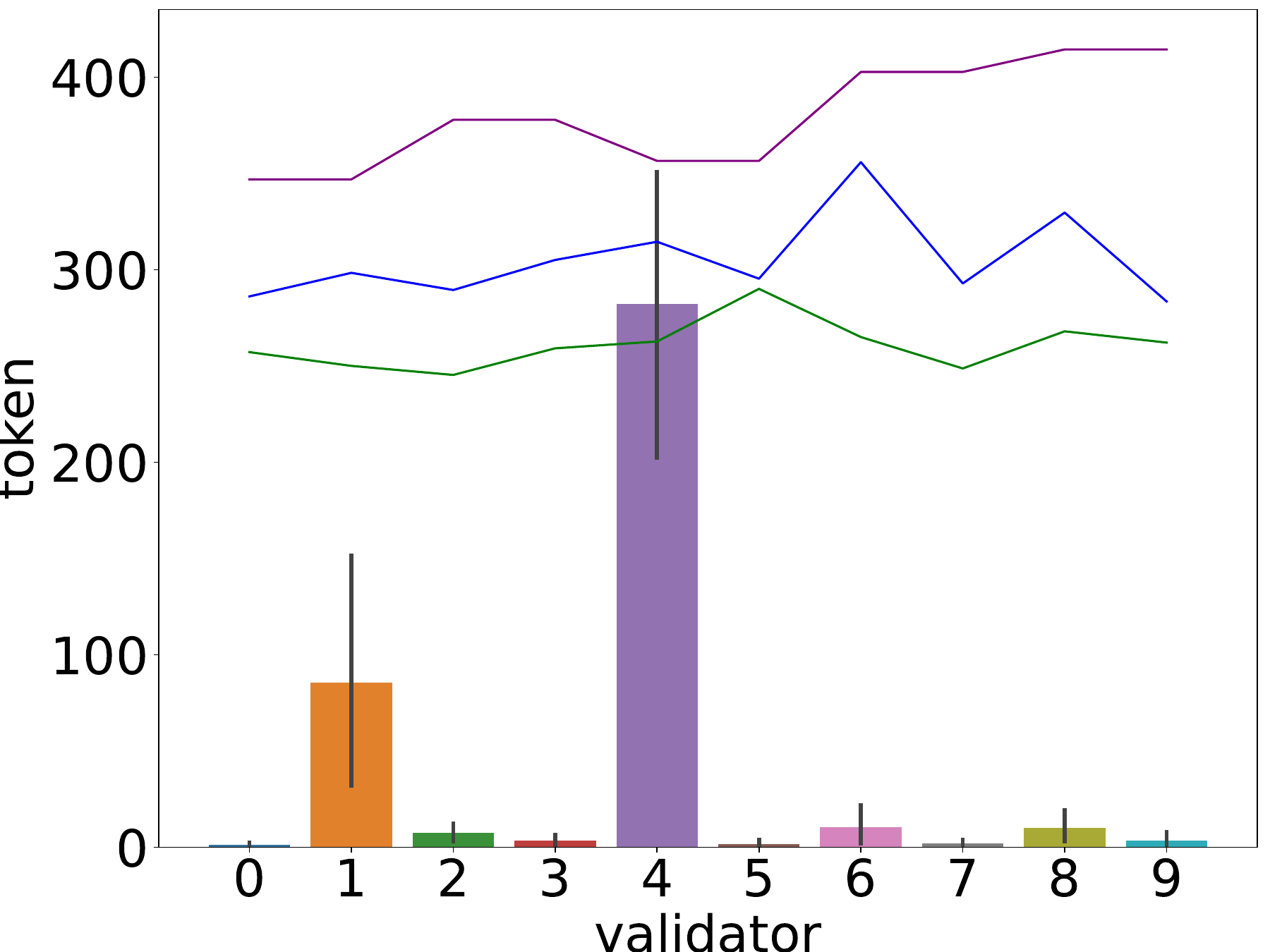}}
\subcaptionbox
{$\epsilon=0.9$ with $2$ round. \label{fig:e9r2}}{\includegraphics[width=4.3cm]{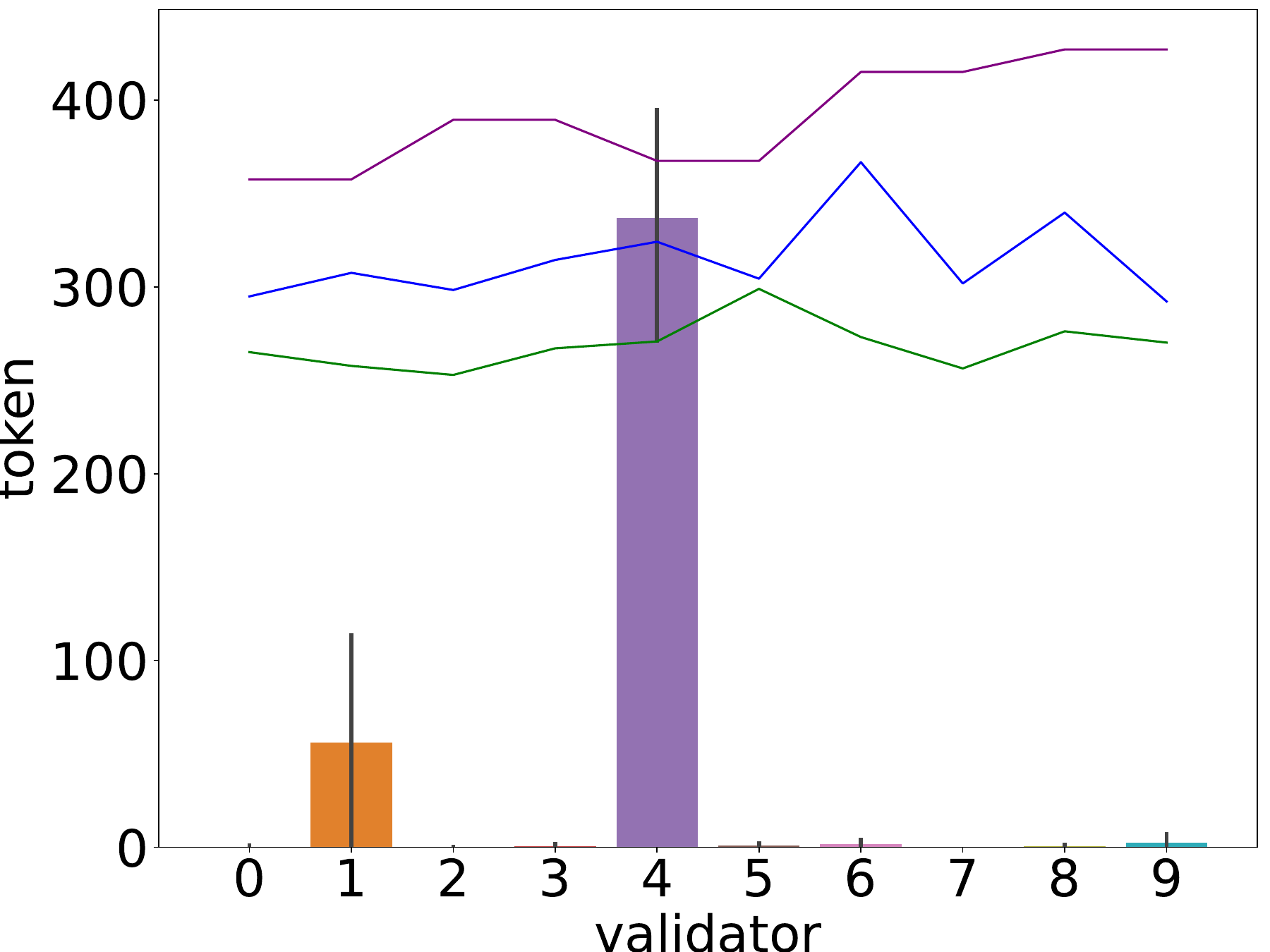}}
\subcaptionbox
{$\epsilon=1$ with $2$ round. \label{fig:e10r2}}{\includegraphics[width=4.3cm]{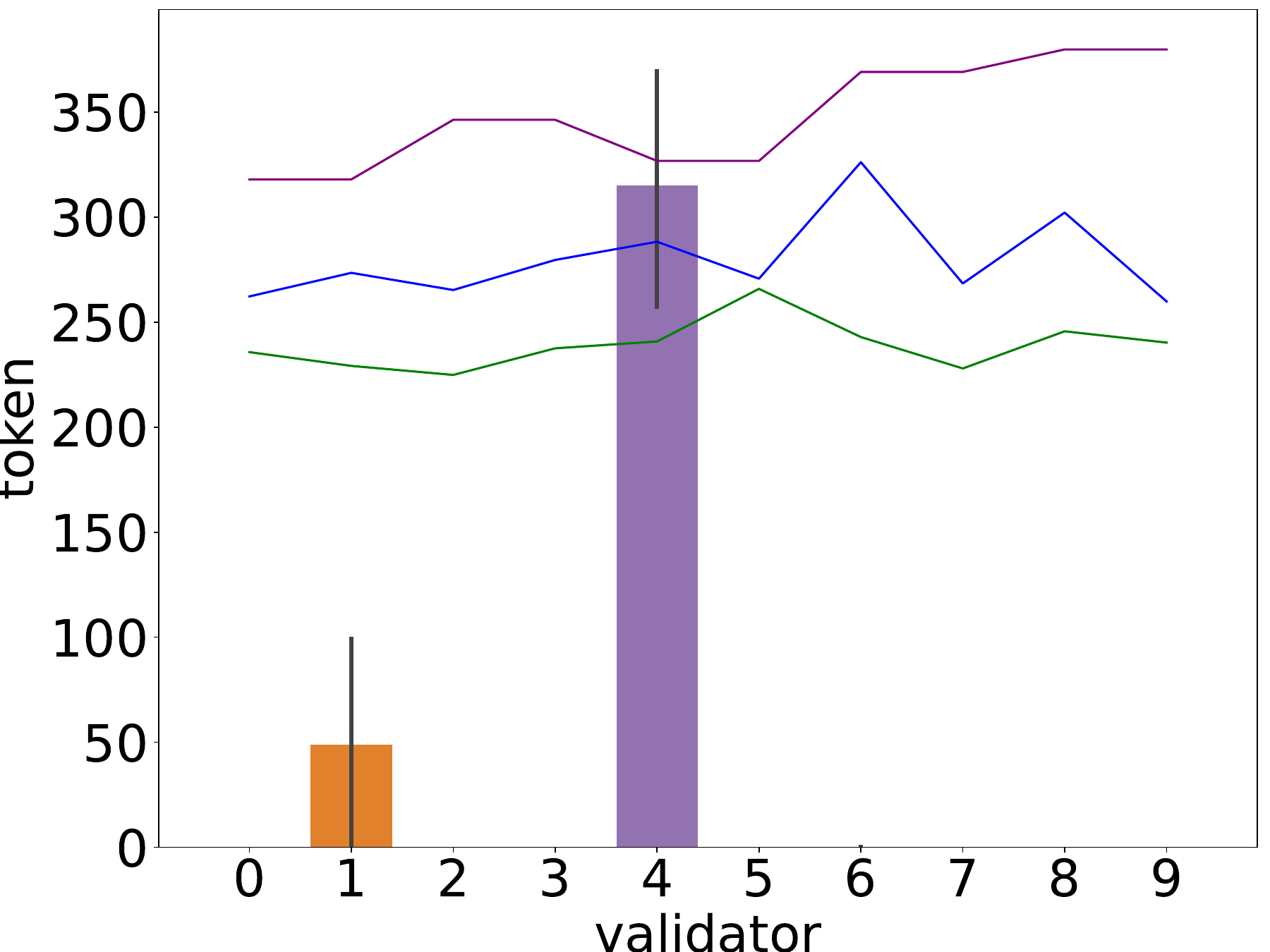}}
 \caption{\textit{Bars}: the average values of each validator's received delegation tokens in $20$ simulation instances, varying $\epsilon$ and $RL$ (i.e., the round limit). \textit{Lines}: \textbf{\textcolor{purple}{---}} validator integrities $p_j$, \textbf{\textcolor{blue}{---}} validator evidences $\Pr(b_j=1)$, \textbf{\textcolor{darkgreen}{---}} validator commission rates $c_j$.}
 \label{fig:simu}
\end{figure*}
\subsubsection{Commission-free VSG}\label{subsec-cfree}\hfill
\medskip

Lastly, we consider the subclass of commission-free VSG where the validator's commission rate is $0$.
We show that if commission-free VSG satisfies certain conditions, there always exists an NE.

\begin{thm}\label{thm:NEcf}
In a commission-free VSG, a NE is always guaranteed to exist, if, for each user $a_i\in A$, their budget satisfies $b_i \ge \frac{(n-1)r}{n^2(1-T)}\ge 1$, where $T=\frac{q^np^*}{q^np^*+\bar{q}^n(1-p^*)}$ such that $p^*$ is the maximal integrity among all validators.
\end{thm}

\begin{prf}[Thm.\ref{thm:NEcf}]
We prove this theorem by showing that strategy profile $(\mathbf{b},\mathbf{t})$ such that for each user $a_i\in A$, $b_i=v^*$ 
($v^*\in V$ is a validator with the maximal integrity, i.e., $v^*\in arg\max_{v_j\in V}p_j$) and $t_i=\frac{(n-1)r}{n^2(1-T)}$ is a NE.
Intuitively, $(\mathbf{b},\mathbf{t})$ is such a strategy profile that each user delegates the above number of tokens to the same validator who has the maximal integrity.
We illustrate that each user $a_i$ cannot obtain a higher utility by unilaterally altering their strategy in three exhaustive ways:  \textbf{(i)} to abstain;  \textbf{(ii)} to only alter their token strategy; and  \textbf{(iii)} to delegate to another validator.

\noindent\textbf{\textit{Case-}(i).}
Taking strategy $(\mathbf{d}, \mathbf{t})$, $a_i$ obtains utility
\begin{align*}
u_i(\mathbf{d}, \mathbf{t})=\frac{r}{n^2}>0.
\end{align*}
Thus, $a_i$ has no incentive to deviate from $(\mathbf{d},\mathbf{t})$ to abstaining otherwise $a_i$ obtains a lower utility of $0$.

\noindent\textbf{\textit{Case-}(ii).}
We can reason this case by the same argument in the proof of \underline{Thm.\ref{thm:NEsingle}}.

\noindent\textbf{\textit{Case-}(iii).}
Assume that $a_i$ delegates to validator $v_{j'}$ (either $v_{j'}=v_j$ or $v_{j'}\not= v_j$).
We obtain that, the maximal utility that $a_i$ can achieve by taking the optimal token strategy $t'$ computed by \underline{Eq.\ref{eq:optimaltoken}} is
\begin{align*}
u^*_i=\left(\sqrt{\frac{(1-T_{ij'})(n-1)T^{(n-1)}}{T_{ij'}}}-\sqrt{r}\right)^2,
\end{align*}
where $T^{(n-1)}$ is, based on strategy profile $((\mathbf{d}_{-i},d_i=v_{j'}),(\mathbf{t}_{-i},t'))$, the trust of all users other than $a_i$ on $v^*$, and $T_{ij'}$ is the trust of $a_i$ on $v_{j'}$.

\noindent Compare two cases: (1) $v_{j'}=v_j$, and (2) $v_{j'}\not= v_j$.
From (1) to (2), we have that $T_{ij'}$ decreases by \underline{Lm.\ref{le:monotone}} and \underline{Lm.\ref{le:tobettervali}}, since $p_{j'}\le p_{j}$.
We let $X=\sqrt{\frac{(1-T_{ij'})(n-1)T^{(n-1)}}{T_{ij'}}}$, and thus, we have that $X$ increases from (1) to (2), since $T_{ij'}$ decreases, and $frac{T^{(n-1)}}{T_{ij'}}$ increases from $1$ to larger than $1$ from (1) to (2) due to $T^{(n-1)}>T_{ij'}$ by \underline{Lm.\ref{le:monotone}}.

Since $u_i\sqrt{\frac{(1-T_{ij'})(n-1)T^{(n-1)}}{T_{ij'}}}>0$, otherwise $a_i$ would prefer to abstain, we have that 
\begin{align*}
\frac{rT_{ij'}}{(n-1)T^{(n-1)}t}-(1-T_{ij'})>0.
\end{align*}
By condition $t>1$, we further obtain that $X<\sqrt{r}$.
Therefore, we have that $u^*_i$ decrease from (1) to (2), which completes the proof.
\qed
\end{prf}

We can observe that, by \underline{Thm.\ref{thm:NEhomov}} and \underline{Thm.\ref{thm:NEcf}}, in a homogeneous VSG or a commission-free VSG, when users' budgets are high enough, the strategies of users may converge to such a structure that all users delegate to the same validator, who has the maximal integrity in the commission-free VSG, and users delegate with the same token strategy.

We configure that this is a reasonable result in VSGs.
In a market, there should exist an optimal validator who has a good trade-off between their reputation and commission.
Then, users would tend to delegate to this validator, and this validator is more trusted as more users delegate to them.

\section{Experiment}
\label{sec-experiment}

Relaxing the assumptions in the above theoretical analysis, we now empirically investigate users' behavior by computer simulations.
In the empirical study, we consider users can only take token strategies in a discrete space, which coincides with the practice.
The assumptions on users' accuracies and errors are relaxed, such that
misled users (i.e., $\exists a_i\in A$ and $\exists v_j\in V$, $q_{ij} < 0.5$ or $\bar{q}_{ij} > 0.5$) exist.
Additionally, we also consider that validators' integrities are not revealed perfectly, i.e., $\exists v_j\in V$, the evidence quality (recall Equation~\ref{eq:trust}) of $v_j$ satisfies $q_j<1$.

With the above practical settings, we use a noisy model to simulate that a set of users, in turn, choose their {\it best response strategy} against the current strategy profile, in a random \textit{round-robin} order.

\begin{defi}[Best Response]\label{defi-bestresponse}
Given a VSG and a strategy profile $(\mathbf{d},\mathbf{t})$, for each user $a_i\in A$, their best response strategy against $(\mathbf{d},\mathbf{t})$ is the strategy $(d'_i,t'_i)$ by which $a_i$ can obtain the highest utility assuming that the other users do not change their strategies in $(\mathbf{d},\mathbf{t})$.
Formally, given a strategy profile $(\mathbf{d},\mathbf{t})$, any strategy $(d'_i,t'_i)\in arg\max_{(d_i,t_i)\in \Sigma_i}u_i(((\mathbf{d}_{-i},d_i),(\mathbf{t}_{-i},t_i)))$ is $a_i$'s best response strategy.
\end{defi}
We design a so-called $\epsilon$-Greedy Best Response Dynamics (\underline{Alg.\ref{algo:GBRD}}, with abbreviation GBRD) to simulate users' behavior.
In GBRD, we initiate the strategy profile by assigning a validator to each user at random, with a randomly sampled integer between $0$ and $1/(1+c_j)$ of their budget as the token strategy, where $c_j$ is the commission rate of each user's randomly assigned validator.
The initiated strategy profile is denoted as $(\mathbf{d}^0,\mathbf{t}^0)$.
Then, in each round, following a randomly generated order $\sigma$ (i.e., a random permutation of all users in $A$), against the current strategy profile, each user in turn chooses their best response strategy by an $\epsilon$-greedy manner.
That is, given a strategy profile $(\mathbf{d},\mathbf{t})$ and a user $a_i\in A$, let $R_i^{\text{best}}(\mathbf{d},\mathbf{t})=arg\max_{(d',t')\in \Sigma_i}u_i((\mathbf{d}_{-i},d'),(\mathbf{t}_{-i},t'))$ be the set of best response strategies of $a_i$, and $R_i^{\text{better}}(\mathbf{d},\mathbf{t})=\{(d',t')\in \Sigma_i\mid u_i((\mathbf{d}_{-i},d'),(\mathbf{t}_{-i},t'))>u_i(\mathbf{d},\mathbf{t})\}\setminus R_i^{\text{best}}$ be the set of strategies that can improve $a_i$'s utility against $(\mathbf{d},\mathbf{t})$ except for those in $R_i^{\text{best}}$.
Then, in GBRD, at each round, each user uniformly at random chooses a strategy from $R^{\text{best}}_i$ with probability $\epsilon\in [0,1]$, and uniformly at random chooses one from $R^{\text{better}}_i$ with probability $1-\epsilon$, in the turn of $\sigma$.
After a round, if for each user, the change of their utility is less than a ratio $\theta$ of their utility of the last round, or the number of rounds reaches the limit $RL$, the algorithm terminates and returns the strategy profile output by the last round.

\begin{algorithm}[h]
\begin{description}
\item[Initialization:] $\sigma$, $\epsilon$, $s=0$, $(\mathbf{d}^{0},\mathbf{t}^{0})$, $RL$, $\theta$.
\item[Iteration:]\hfill\\
Round $s$:
\begin{itemize}
\item \textcolor{black}{\textit{Step1}}: $k=1$, and $(\mathbf{d}^{s+1},\mathbf{t}^{s+1})=(\mathbf{d}^s,\mathbf{t}^t)$.\\
\item \textcolor{black}{\textit{Step2}}: For $a=\sigma(k)$, compute $u_{a}(((\mathbf{d}^{s+1}_{-a},d_k),(\mathbf{t}^{s+1}_{-a},t_k)))$ for all $(d_k,t_k)\in \Sigma_{a}$.\\
\item \textcolor{black}{\textit{Step3}}: With probability of $\epsilon$, $a$ samples a strategy from $R^{\text{best}}(\mathbf{d}^{s+1},\mathbf{t}^{s+1})$ uniformly at random, and with probability of $(1-\epsilon)$, $a$ samples a strategy from $R^{\text{better}}(\mathbf{d}^s,\mathbf{t}^s)$. Replace $(d^{s+1}_a,t^{s+1}_a)$ by the sampled strategy.\\
\item \textcolor{black}{\textit{Step4}}: If $k<n$, $k\leftarrow k+1$ and go to \textcolor{black}{\textit{Step2}}, else to \textcolor{black}{\textit{Step5}}.\\
\item \textcolor{black}{\textit{Step5}}: If $|u_i(\mathbf{d}^{s+1},\mathbf{t}^{s+1})-u_i(\mathbf{d}^s,\mathbf{t}^s)|\le \theta$ or $s+1=RL$, terminate, else $s\leftarrow s+1$ and go back to \textcolor{black}{\textit{Step1}}.
\end{itemize}
\item[Return:] $(\mathbf{d}^{s+1},\mathbf{t}^{s+1})$
\end{description}
\caption{$\epsilon$-Greedy Best Response Dynamics}
\label{algo:GBRD}
\end{algorithm}

\subsection{Experiment Settings}\label{subsec-setting}
We conduct simulations with $10$ validators and $200$ users.
The integrity and evidence quality of each validator, and the accuracy and error of each user are randomly generated by Gaussian distributions with a standard deviation of $0.1$, and the means of integrity, evidence quality, accuracy, and error are $0.7$, $0.8$, $0.6$, $0.5$, respectively.
Especially, each generated value of integrity, evidence quality, and accuracy is forced in the interval $[0.5, 1]$.
The budget of each user is generated by a Gaussian distribution of $N(70, 15)$, and the profit is $r=30$.
The commission rate of each validator $v_j\in V$ is generated as $c_j=(p_j-0.5)/3+\delta$, where $\delta$ is a Guassian error following $N(0,0.01)$, and $c_j$ is forced positive.
Each commission rate is in the range $[0,0.2]$ with high probability.
Lastly, $\theta$ is set as $0.01$.

We run all simulation instances under the same initiative setting of validator integrities, evidence qualities and commission rates, and user accuracies, errors, and budgets.
For each parameter, we run 20 simulation instances.
The following results are the average of values output by 20 instances.

We conduct the experiment in Python 3.9, and run the simulation on a MacBook Pro with an Apple M1 Pro chip.

\subsection{Results}\label{subsec-result}
We first observe that, according to the noisy model, information error exists in the market: validators' integrities (purple line) and evidence (blue line) do not have the same trend, suggesting that a validator with high integrity may not necessarily have high evidence (i.e., a good reputation).

Our results also show that almost all users delegate: averagely more than $99\%$ users delegate in each parameter setting.
Among those who delegate, averagely, each user only uses less than $10\%$ of their tokens, with a range from $2.7\%$ to $9.9\%$ in all parameter settings.
Users delegate fewer tokens when they are able to specify their optimal strategies more accurately, i.e., corresponding to a higher $\epsilon$ or a higher round number $RL$.
We configure that it is because we use a small profit $r=30$, which leads the users to choose a low level of token strategies such that the users would not lose too much due to the commission and the risk of trust.

Generally, observe that in all figures in \underline{Fig.\ref{fig:simu}}, users' delegation tokens tend to concentrate on a small part of validators. However, most of the validators receiving a large amount of tokens are not among those with high integrity or high evidence, except for validator $6$ in results of $\epsilon=0.7, 0.8$ and $0.9$ (i.e., Fig.\underline{\ref{fig:e7r1}}, \underline{\ref{fig:e8r1}}, \underline{\ref{fig:e9r1}}).
Instead, tokens are concentrated on validators with a good balance of high reputation and low commission rate, e.g., validators $4$ and $1$.

Fig.\underline{\ref{fig:e7r1}}, \underline{\ref{fig:e8r1}}, \underline{\ref{fig:e9r1}} and \underline{\ref{fig:e10r1}} illustrate to which validators the users delegate their tokens when they in turn noisily choose their best response strategy only once ($RL=1$ in \underline{Alg.\ref{algo:GBRD}}), by varying the noise indicator $\epsilon$ from $0.7$ to $1$.
Observe that as $\epsilon$ becomes higher, tokens are more concentrated, especially on validators $4$ and a relatively small amount on validator $1$.
This coincides with our configuration that, as $\epsilon$ becomes higher, users are more accurate in delegating to the best validator with the optimal token strategy with respect to the utility defined in Equation~\ref{eq:utility}.
Though a large amount of tokens are also delegated to validators $1$ and $6$ in smaller $\epsilon$'s (Fig.\underline{\ref{fig:e7r1}}, \underline{\ref{fig:e8r1}}, \underline{\ref{fig:e9r1}}), those amounts gradually decrease as $\epsilon$ becomes larger.
We can conclude that when users' rationality is high (i.e., corresponding to higher $\epsilon$), they can better converge to strategy profiles where they delegate to the validators with the best balance of good reputation and low commission.
We can also conclude that, in this randomized simulation instance, validator $4$ should be the one with the best balance of good reputation and low commission for users, and validator $1$ also shows a good trade-off on these two attributes.

A similar trend can be observed in Fig.\underline{\ref{fig:e7r2}}, \underline{\ref{fig:e8r2}}, \underline{\ref{fig:e9r2}}, \underline{\ref{fig:e10r2}}, where users go through $2$ rounds of iteration in \underline{Alg.\ref{algo:GBRD}}, varying $\epsilon$ also from $0.7$ to $1$.
However, we can also observe that for each $\epsilon$, tokens are better concentrated in figures of $2$ rounds (the second row) than figures of $1$ rounds (the first row).
This is because, by each iteration, users can gradually improve their strategy. More users will thereby delegate their tokens to the optimal validators.
This further supports the conclusion that validators $4$ and $1$ are the best choices for users in terms of good reputation and low commission rate.

\section{Conclusion}
This paper analyses the delegation process as regular users choose their validators, offering models to quantify virtual trust in delegatees. Through game-theoretical simulations, we explore user behavior while considering vital factors from active staking services and PoS blockchains. Our findings indicate that users make decisions that lead to a Nash equilibrium by weighing delegation costs, other users' actions, and delegatee reputation. However, this trend heightens the risk of token concentration among a few delegatees.


\smallskip
\noindent\textbf{\textcolor{red}{\dangersign} Open problems for NEXT.} We present several future plans.

\smallskip
\noindent\ding{172} As this work majorly focuses on the behaviors of normal users (stakers), we could extend our game theoretical model by incorporating validators as players. 
Validators may choose a strategy between staying in or leaving the market, based on their potential profit decided by extrinsic factors, such as their cost to run a client and their accrual of delegation tokens. Then, validators with lower integrity might be easily motivated by their potential profit to leave the market. This brings the users more concerns on judging validators' leaving risk than only considering the trust in validators' intrinsic motivation. Our configuration is that this setting may prevent users from concentrating their tokens on a small number of validators, such that the distribution of delegation tokens is more equal among validators.

\smallskip
\noindent\ding{173} We initially model the subjective parameters, such as a user's accuracy, as static within the framework of the game. However, for a more realistic representation, we can introduce a dynamic learning process to account for users' potential growth and accelerated learning during each evolutionary phase. This extension would involve incorporating a dynamic accuracy parameter into the model to better align with practical scenarios.

\smallskip
\noindent\ding{174} As an extended part of experiments, we plan to perform a comparative analysis that involves evaluating the simulated data generated by our model alongside real-world data collected from publicly available sources related to stakers and staking providers. 


\newpage

\normalem
\bibliographystyle{unsrt}
\bibliography{bib.bib} 

\appendix



\section{Additional Experiment Results}
\label{apx-exp}

\begin{figure*}[!hbt]
\centering
\subcaptionbox
{$\epsilon=0.7$ with $3$ round. \label{fig:e7r3}}{\includegraphics[width=4.3cm]{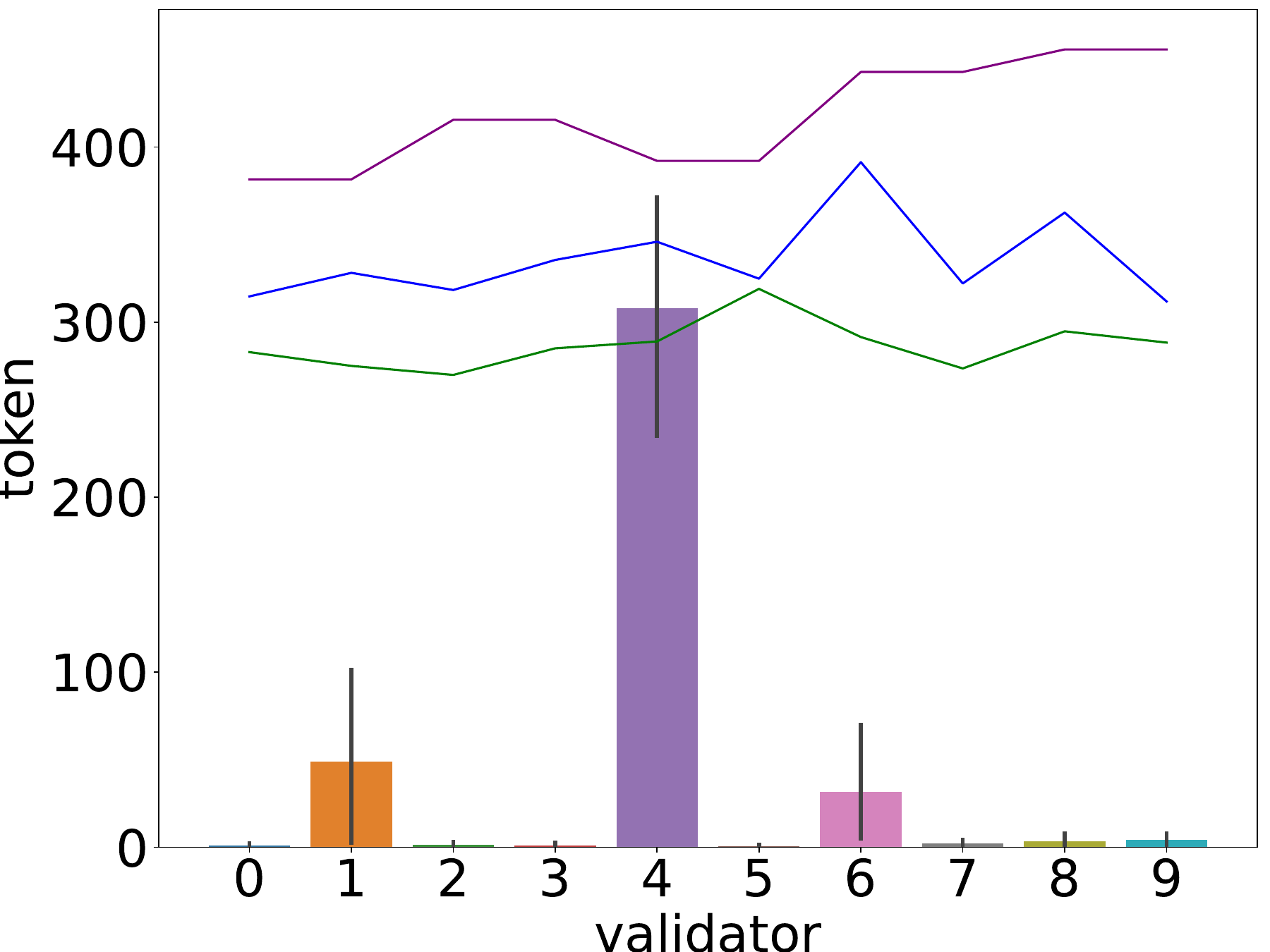}}
\subcaptionbox
{$\epsilon=0.8$ with $3$ round. \label{fig:e8r3}}{\includegraphics[width=4.3cm]{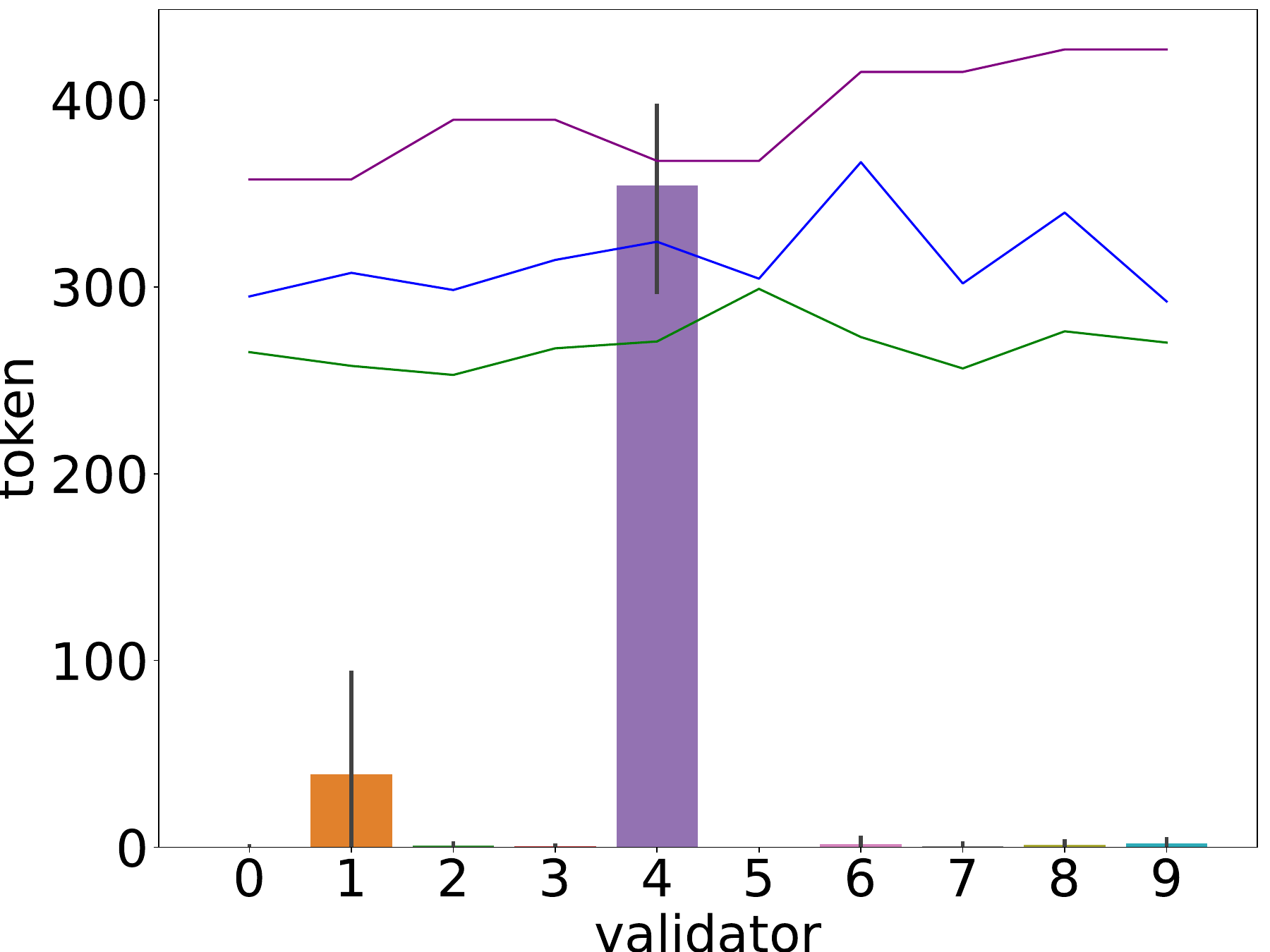}}
\subcaptionbox
{$\epsilon=0.9$ with $3$ round. \label{fig:e9r3}}{\includegraphics[width=4.3cm]{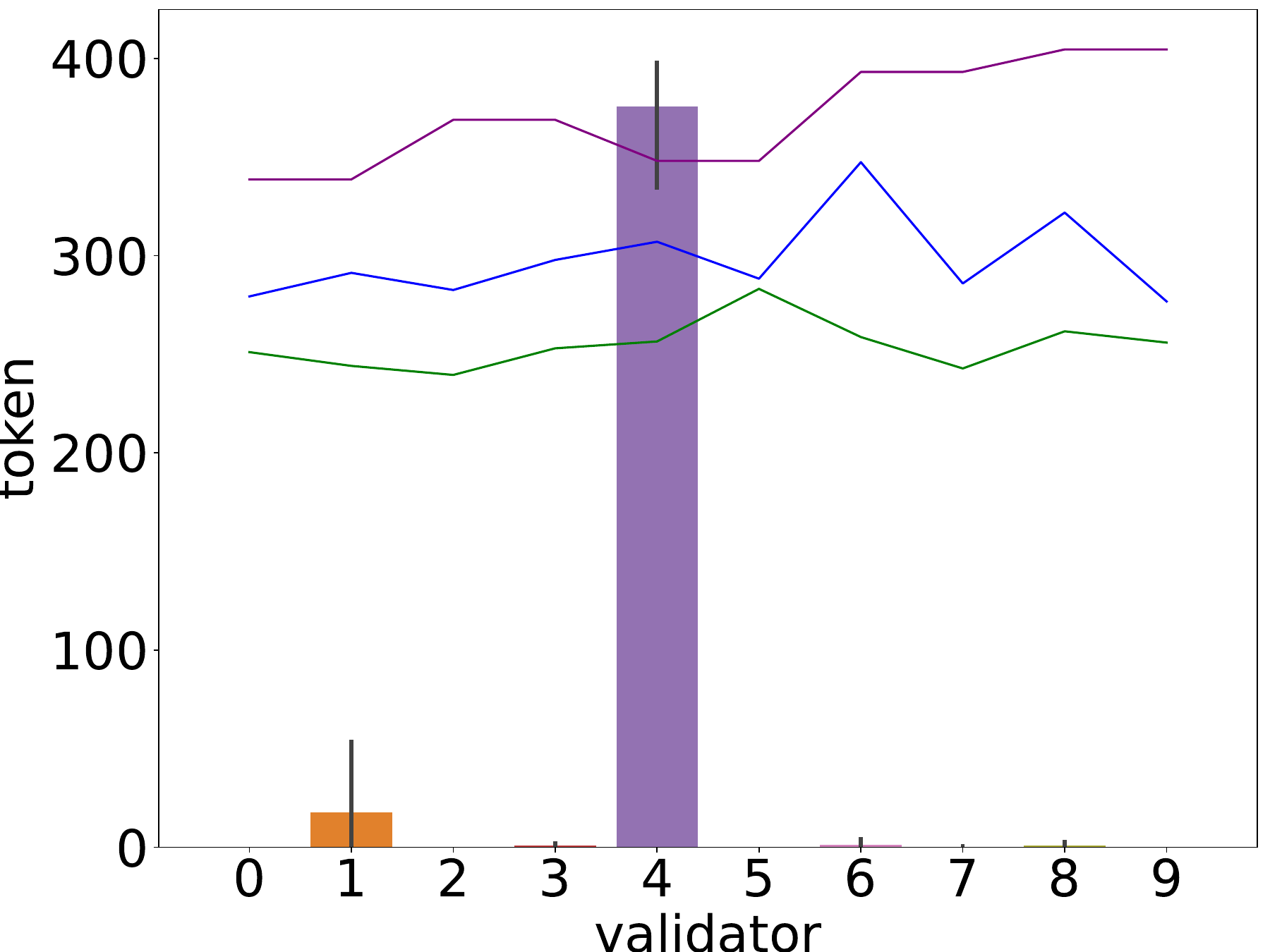}}
\subcaptionbox
{$\epsilon=1$ with $3$ round. \label{fig:e10r3}}{\includegraphics[width=4.3cm]{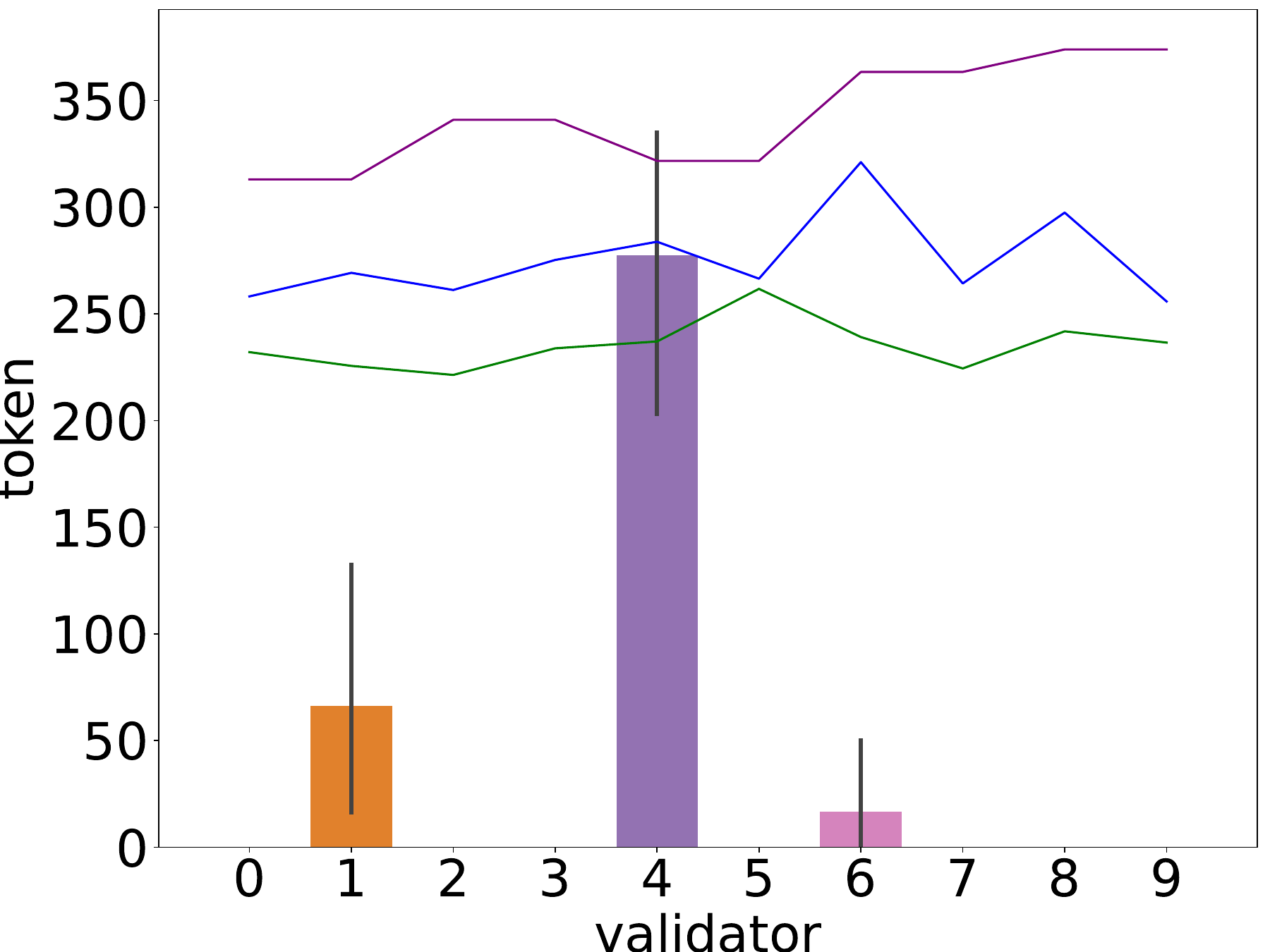}}
 \caption{\textit{Bars}: the average values of each validator's received delegation tokens in $20$ simulation instances, varying $\epsilon$ and fixing $RL=3$ (i.e., the round limit). \textit{Lines}: \textbf{\textcolor{purple}{---}} validator integrities $p_j$, \textbf{\textcolor{blue}{---}} validator evidences $\Pr(b_j=1)$, \textbf{\textcolor{darkgreen}{---}} validator commission rates $c_j$.}
 \label{fig:simu_app}
 \vspace{-0.2cm}
\end{figure*}

\begin{figure*}[!hbt]
\centering
\subcaptionbox
{$\epsilon=0.7$ with $1$ round. \label{fig:token_r1e7}}{\includegraphics[width=18cm]{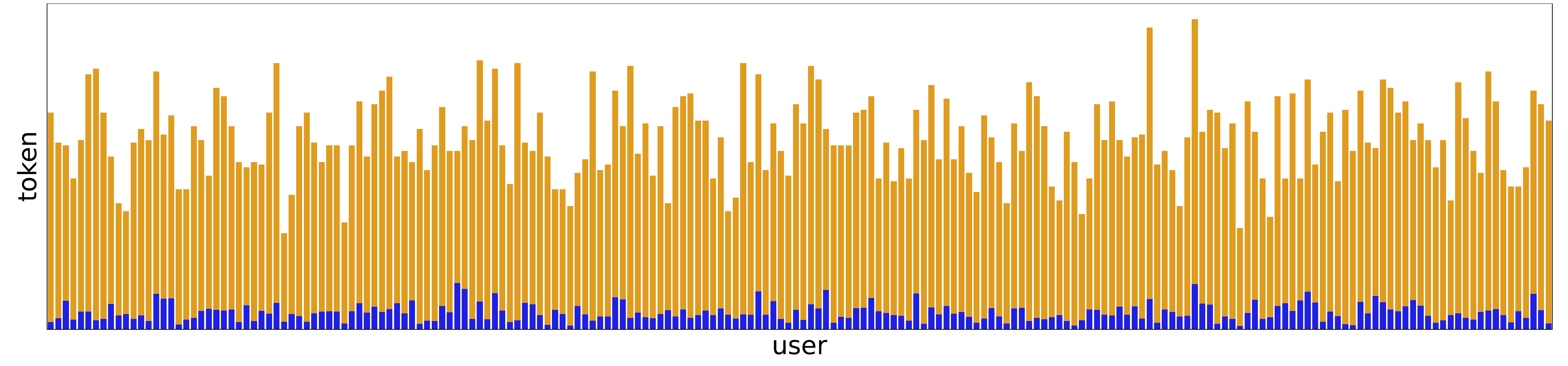}}
\subcaptionbox
{$\epsilon=1$ with $1$ round. \label{fig:token_r1e10}}{\includegraphics[width=18cm]{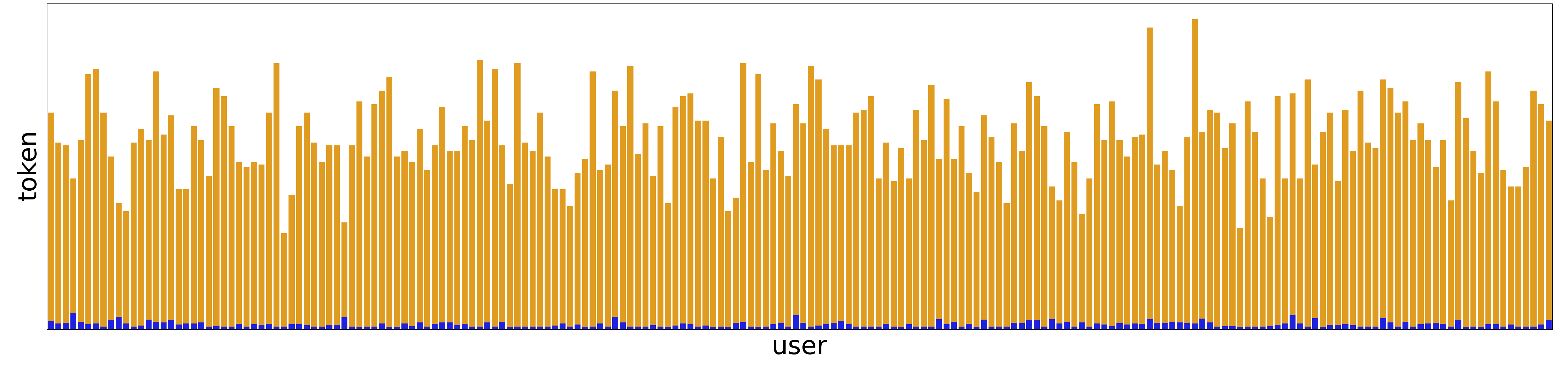}}
\subcaptionbox
{$\epsilon=0.7$ with $3$ round. \label{fig:token_r3e7}}{\includegraphics[width=18cm]{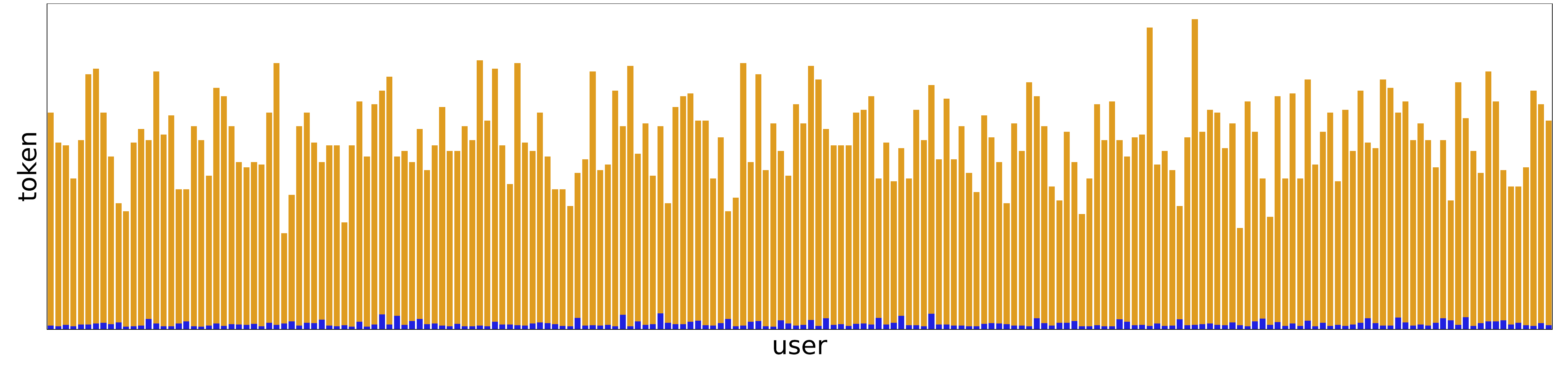}}
 \caption{Users' token usage. \colorbox{orange}{\textcolor{orange}{-}}: users' budgets, \colorbox{blue}{\textcolor{blue}{-}}: users' average number of delegating tokens. }
 \label{fig:token_app}
\end{figure*}

\begin{figure*}[!hbt]
\centering
\subcaptionbox
{$p_j=0.6$, $(q,\bar{q})=(0.6,0.7)$. \label{fig:p6q6}}{\includegraphics[width=4.3cm]{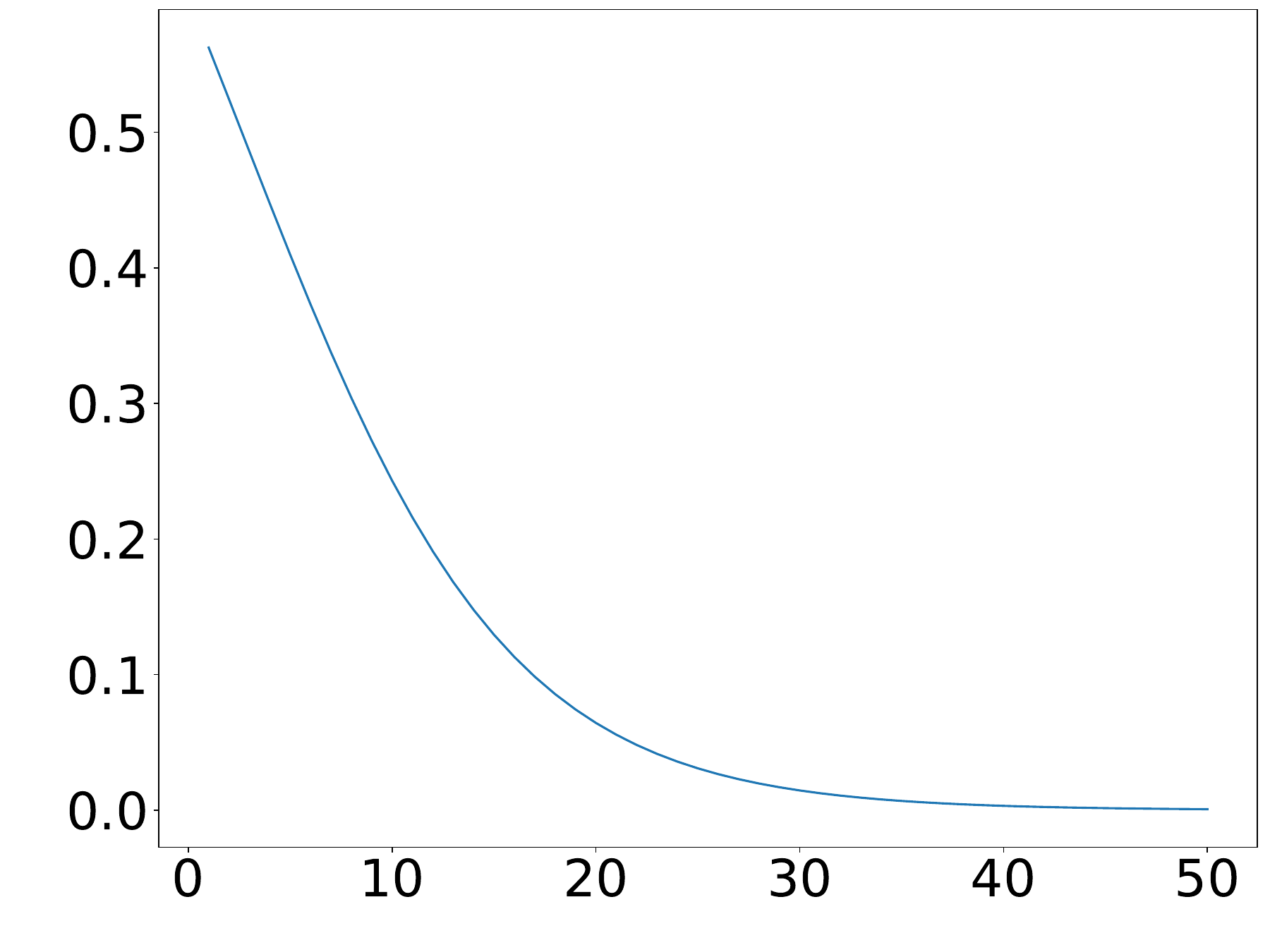}}
\subcaptionbox
{$p_j=0.6$, $(q,\bar{q})=(0.7,0.6)$. \label{fig:e8r2}}{\includegraphics[width=4.3cm]{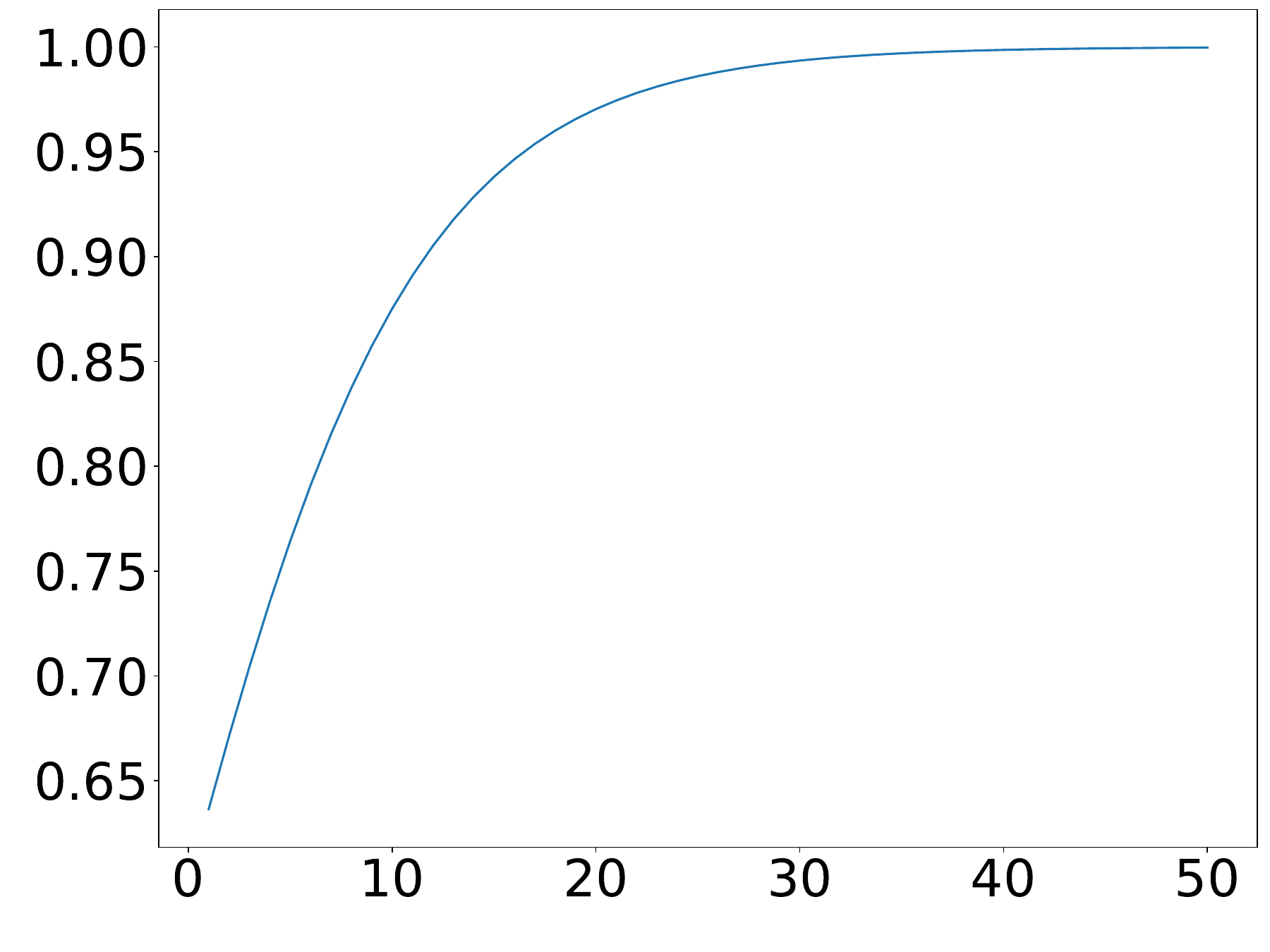}}
\subcaptionbox
{$p_j=0.9$, $(q,\bar{q})=(0.6,0.7)$. \label{fig:e9r2}}{\includegraphics[width=4.3cm]{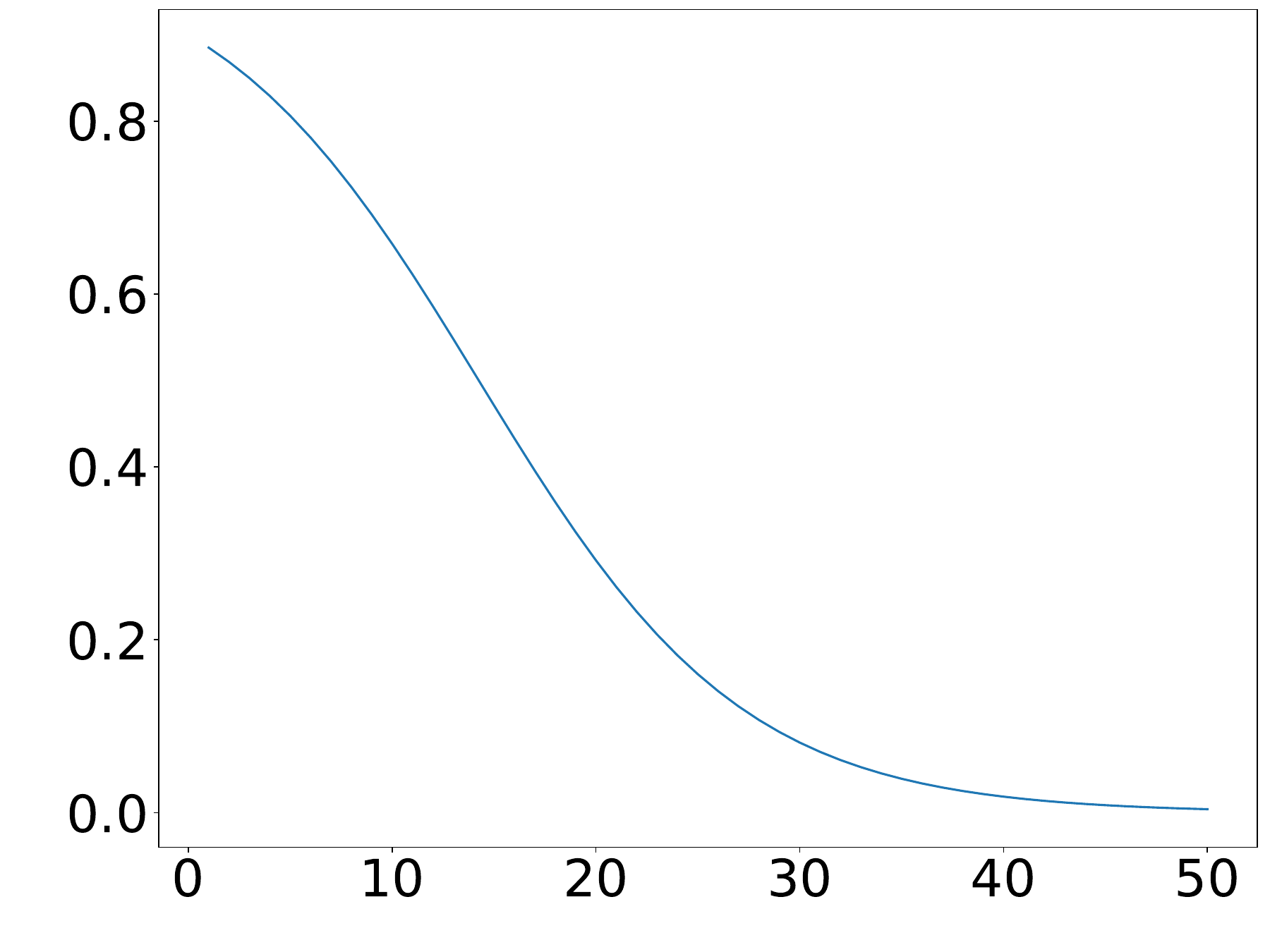}}
\subcaptionbox
{$p_j=0.9$, $(q,\bar{q})=(0.7,0.6)$. \label{fig:e10r2}}{\includegraphics[width=4.3cm]{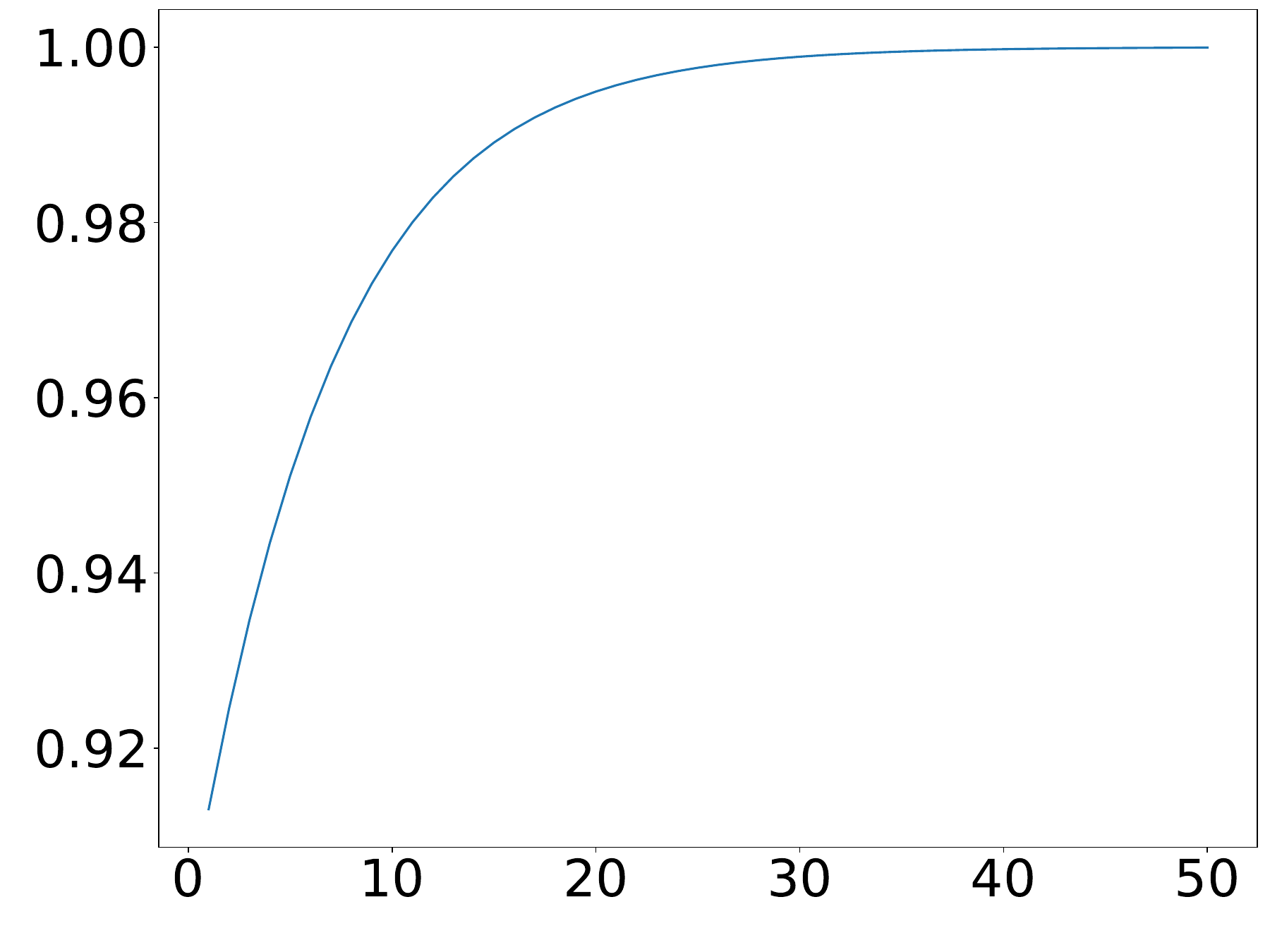}}

\vspace{-0.1in}
 \caption{Users' trust on a validator $v_j$, by varying integrity $p_j$ and users' accuracy and error $(q,\bar{q})$. The x-axis denotes the number of users delegate to $v_j$.}
 \label{fig:trust_app}
 \vspace{-0.2cm}
\end{figure*}

We show more details corresponding to the description in \underline{Sec.\ref{subsec-result}}.
\underline{Fig.\ref{fig:simu_app}} provides the results of running \underline{Alg.\ref{algo:GBRD}} more rounds than results shown in \underline{Fig.\ref{fig:simu}}.
\underline{Fig.\ref{fig:token_app}} shows the average ratio of users' delegation token numbers to their budgets.
\underline{Fig.\ref{fig:trust_app}} shows users' trust on a validator by varying the number of homogeneous delegators, in different settings of validator integrity and user accuracy and error.

\section{Staking Services}
\label{apx-stake}

Following \underline{Sec.\ref{subsec-stake}}, we further present a summary of custodial staking services offered by various platforms (majorly CEXes, \underline{Tab.\ref{tab:staking-custodial}}). These platforms vary in terms of user entry requirements, supported proofs, unstaking periods, and maximum staking rates.

\begin{table}[!hbt]
\caption{Staking services (\textit{custodial})}\label{tab:staking-custodial}
\renewcommand\arraystretch{1.1}
\begin{center}
\resizebox{1\linewidth}{!}{
\begin{tabular}{c|cccccc|cc} 

         \multicolumn{1}{c}{\cellcolor{gray!10}\textbf{}} & 
         \cellcolor{gray!10}\textbf{Mini.} & 
         \cellcolor{gray!10}\textbf{Proofs} &  
         \multicolumn{1}{c}{\cellcolor{gray!10}\textbf{Unst.}}   & 
         \multicolumn{1}{c}{\cellcolor{gray!10}\textbf{M. Shr}}   & 
         \multicolumn{1}{c}{\cellcolor{gray!10}\textbf{Valid.}}   & 
         \multicolumn{1}{c}{\cellcolor{gray!10}\textbf{C. Fee}}   & 
         \cellcolor{gray!10}\textbf{Sprt.}  &
         \cellcolor{gray!10}\textbf{Max.}  \\
        \cmidrule{1-2} 

        \cellcolor{gray!10} \hlhref{https://www.coinbase.com/earn/staking/ethereum}{Coinbase}  &  \cellcolor{gray!10} Any  &  \cellcolor{gray!10} cbETH & \cellcolor{gray!10} T+2 & \cellcolor{gray!10} 14\% & \cellcolor{gray!10}  121k &  \cellcolor{gray!10} 25\% & \cellcolor{gray!10} 9  & \cellcolor{gray!10} 6.12\%   \\
   
        \cellcolor{gray!10} \hlhref{https://www.binance.com/en/eth2}{Binance}  & \cellcolor{gray!10} Any  & \cellcolor{gray!10} BETH &  \cellcolor{gray!10}   T+2  & \cellcolor{gray!10} 4.3\% & \cellcolor{gray!10}  37k & \cellcolor{gray!10}   	10\%  & \cellcolor{gray!10} 200+  & \cellcolor{gray!10} 157.81\% \\
        
        \cellcolor{gray!10} \hlhref{https://www.kraken.com/features/staking-coins/ethereum}{Kraken}  & \cellcolor{gray!10} Any  & \cellcolor{gray!10} ``staked"  & \cellcolor{gray!10} Secs & \cellcolor{gray!10}  3\% & \cellcolor{gray!10}  26k & \cellcolor{gray!10} 15\%  & \cellcolor{gray!10} 15+  & \cellcolor{gray!10}  26\%    \\
        
        \cellcolor{gray!10} \hlhref{https://www.kucoin.com/earn/eth2}{KuCoin}   & \cellcolor{gray!10} Any & \cellcolor{gray!10}ksETH & \cellcolor{gray!10} T+5 & \cellcolor{gray!10} 0.1\% & \cellcolor{gray!10}  2k & \cellcolor{gray!10} - & \cellcolor{gray!10} 40+ &  \cellcolor{gray!10}   180.15\%   \\

        \cellcolor{gray!10} \hlhref{https://support.bake.io/en/articles/8286119-eth-staking-faqs}{Cake DeFi}   & \cellcolor{gray!10} Any  & \cellcolor{gray!10} csETH  & \cellcolor{gray!10} - &  \cellcolor{gray!10} 0.1\% & \cellcolor{gray!10}  2k  & \cellcolor{gray!10} -  &\cellcolor{gray!10} 4  & \cellcolor{gray!10} 12.4\%    \\

        \cellcolor{gray!10} \hlhref{https://crypto.com/university/how-to-stake-ethereum}{Crypto.com}  & \cellcolor{gray!10} $1e^{-8}$  & \cellcolor{gray!10}  ``staked"  & \cellcolor{gray!10}  T+5   & \cellcolor{gray!10} - & \cellcolor{gray!10} - & \cellcolor{gray!10} - & \cellcolor{gray!10} 20+ &  \cellcolor{gray!10} 12\%  \\
        
        \cellcolor{gray!10} \hlhref{https://support.nexo.com/s/article/eth-smart-staking-neth-explained?language=en_US}{Nexo}  & \cellcolor{gray!10} US\$10  & \cellcolor{gray!10} NETH & \cellcolor{gray!10} -  &\cellcolor{gray!10} - & \cellcolor{gray!10} - & \cellcolor{gray!10} 0.20\% & \cellcolor{gray!10}  30+ & \cellcolor{gray!10} 24\%    \\

        \cmidrule{7-8}
        
          \multicolumn{1}{c}{} & \multicolumn{6}{c}{\cellcolor{gray!10}\textbf{Ethereum 2.0}}  &  \multicolumn{2}{c}{\cellcolor{gray!10}\textbf{Other PoS}}   \\

\end{tabular}
}
\end{center}
\end{table}

Then, we also present a comparison of two staking types across various key parameters (\underline{Tab.\ref{tab:staking-comparison}}). Custodial staking relies on a single third party, typically requiring less prior knowledge from users and often not imposing a minimum deposit. Unstaking periods are short due to the providers maintaining a flexible pool of unfrozen tokens for liquidity. However, custodial staking comes with the drawback that users do not have control over their private keys, exposing them to higher delegation risks (e.g., FTX collapse~\cite{fu2022ftx}).

Conversely, non-custodial staking presents contrasting attributes in several dimensions. It involves multiple validators, sets a higher entry threshold for users, and enforces extended unbonding (equiv. unstaking/unfreezing/redemption) periods. Additionally, it comes with increased operational expenses for running a full node. On the plus side, users retain control of their accounts through private keys and benefit from reduced delegation risks.

\begin{table}[!hbt]
\caption{Comparisons for staking types}\label{tab:staking-comparison}
\renewcommand\arraystretch{1.1}
\begin{center}
\resizebox{1\linewidth}{!}{
\begin{tabular}{c|cc} 

         \multicolumn{1}{c}{} &  \multicolumn{1}{c}{\cellcolor{gray!10}\textbf{Custodial}} & \cellcolor{gray!10}\textbf{Non-custodial}   \\
        \cmidrule{1-2} 
        
        \cellcolor{gray!10} \textit{Reliance} & \cellcolor{gray!10} one single third-party  &   \cellcolor{gray!10} multiple validators \\

        \cellcolor{gray!10} \textit{User perception}  & \cellcolor{gray!10} less required & \cellcolor{gray!10}  technical knowledge required \\

        \cellcolor{gray!10} \textit{Mini. deposit}  & \cellcolor{gray!10} mostly no required & \cellcolor{gray!10}  often required \\

        \cellcolor{gray!10} \textit{Unstaking period}  &  \cellcolor{gray!10} flexible (seconds to days) & \cellcolor{gray!10} long (due to the network) \\

        \cellcolor{gray!10} \textit{Maintainance}  & \cellcolor{gray!10} fully delegated no further costs & \cellcolor{gray!10}  need to run a full node \\

        \cellcolor{gray!10} \textit{Account}  &  \cellcolor{gray!10} do not hold private keys ($pk$) & \cellcolor{gray!10} owned by self with $pk$s \\

        \cellcolor{gray!10} \textit{Delegation cost}  &  \cellcolor{gray!10} high (service fees, low APR, etc.) & \cellcolor{gray!10} low (commission fee only) \\
\end{tabular}
}
\end{center}
\end{table}

Additionally, we provide an overview (\underline{Tab.\ref{tab:staking-blockchain}}) of the relationships between non-custodial staking providers and PoS blockchains. 

\smallskip
\noindent\textbf{Staking types}. Lastly, we explain four staking options.

\smallskip
\noindent \textcolor{teal}{$\diamond$} \textit{Solo home staking} is seen as the most impactful, offering full control and rewards to users who are ready to commit at least 32 ETH. This method enhances network decentralization but demands technical know-how and a dedicated setup.

\noindent \textcolor{teal}{$\diamond$} \textit{Staking as a Service} suits those who want to stake 32 ETH but prefer a simpler approach. Users delegate the validation process, though trust in the provider is necessary.

\noindent \textcolor{teal}{$\diamond$} \textit{Pooled staking} is an alternative for users with any amount of ETH. It introduces liquidity tokens, making staking more flexible and accessible. Users keep custody of their assets but must be aware that these solutions are third-party creations.

\noindent \textcolor{teal}{$\diamond$} \textit{CEXes} are the least involved option, offering minimal oversight and effort for stakers uncomfortable with self-custody. However, they consolidate large ETH pools, posing centralization risks.

\section{Related Work}
\label{apx-rw}

\noindent\textbf{Ethereum evolution.} Ethereum has been developed for years. The 1.0 version (\textit{Frontier}) was launched in 2015 and introduced the concept of smart contracts to enable decentralized applications (DApps). Subsequently, a series of upgrades~\cite{ethereumhistory} was added to Ethereum to improve its scalability, security, and functionality. Notable milestones include \textit{Homestead} (block index $\mathsf{\#1,150,000}$), \textit{Byzantium} ($\mathsf{\#4,370,000}$), \textit{Constantinople} ($\mathsf{\#7,280,000}$), \textit{Istanbul} ($\mathsf{\#9,069,000}$). The 2.0 version (\textit{Serenity}, the focus of this paper) is a massive upgrade to the Ethereum blockchain that will bring about many transitions including PoW to PoS, EVM to eWASM, and rollups integration. The transformation commenced with the establishment of the parallel Beacon Chain (Dec 1, 2020), and has since witnessed a sequence of significant updates, including \textit{Berlin} ($\mathsf{\#12,965,000}$), \textit{London} ($\mathsf{\#12,965,000}$), \textit{Paris} (a.k.a., \textit{Merge}, $\mathsf{\#15,537,394}$), and \textit{Shanghai} ($\mathsf{\#17,034,870}$). Furthermore, Ethereum's roadmap outlined a series of near-future updates~\cite{ethereumaa}, including \textit{Merge} (PoW to PoS), \textit{Surge} (rollups~\cite{thibault2022blockchain}, data sharding~\cite{wang2019sok}), \textit{Scourge} (PBS~\cite{heimbach2023ethereum}, MEV protection~\cite{yang2022sok}), \textit{Verge} (Verkle tree~\cite{kuszmaul2019verkle}\cite{verkletree}), \textit{Purge} (protocol simplification) and \textit{Splurge} (account abstraction~\cite{wang2023account}, EIP-1559~\cite{liu2022empirical}).

\smallskip
\noindent\textbf{PoS consensus.} The first instance of PoS adoption in blockchain occurred with PPcoin~\cite{king2012ppcoin}, pioneering efforts to replace PoW with PoS within the Bitcoin ecosystem. This motivates a series of new constructions in different aspects, including investigations into potential attacks~\cite{gavzi2018stake,azouvi2020winkle,schwarz2022three,neu2022two}, advancements in provable secure constructions~\cite{kiayias2017ouroboros,david2018ouroboros}, and improvements in properties covering security~\cite{azouvi2022pikachu,tas2023bitcoin}, privacy~\cite{kerber2019ouroboros,wu2023improving}, availability~\cite{neu2021ebb,badertscher2018ouroboros}, consistency~\cite{kiayias2020consistency,blum2020combinatorics}, finality~\cite{d2023simple,stewart2020grandpa}, dynamicity~\cite{d2023recent}, decentralization~\cite{kwon2019impossibility} and fairness~\cite{jh2023fairpos,saad2021pos}. Recognizing the wide-reaching potential of PoS, Ethereum also initiated its incorporation of PoS into its core development by proposing a series of guiding principles and protocols~\cite{buterin2017casper,buterin2020combining,ethpos,moindrot2017proof} (details in \underline{Sec.\ref{sec-background}}).

\smallskip
\noindent\textbf{PoS variants.} 
PoS is adaptable and has given rise to various variants. Delegated Proof of Stake (DPoS)~\cite{saad2020dpos}, adopted by EOS, TRON and Steem, enables users to stake tokens without becoming validators. Validators have the flexibility to adjust the rewards they share with their delegators as an incentive. Nominated Proof of Stake (NPoS) is a consensus model developed by Polkadot~\cite{polkadotNPoS,Abbas2022polkadot} that shares many similarities with DPoS. One significant difference is that if a nominator (delegator) stakes behind a malicious validator, they may also risk losing their stake. Proof of Activity~\cite{bentov2014proof} is a hybrid consensus protocol that integrates elements of both PoW and PoS. Participants have the capability to engage in both mining and staking activities for block validation. 
Proof of Staked Authority (PoSA), implemented by BNB Smart Chain~\cite{bscposa}, combines Proof of Authority (PoA)~\cite{wang2022exploring} (e.g., OpenEthereum, Substrate) and PoS. Validators in this model take turns to forge blocks. A group of 21 active validators is eligible to participate, selected based on the amount of BNB they stake or have delegated behind them.

\smallskip
\noindent\textbf{Staking analyses.} Grandjean~\cite{grandjean2023ethereum} conducted an analysis of the decentralization of staking power within Ethereum's beacon chain, highlighting the centralized distribution of validators' influence, primarily held by a small number of large entities. A similar study by He et al.~\cite{he2020staking} also identified stake concentration and established the existence of a stable equilibrium where no staking pool has an incentive to deviate. John et al.~\cite{john2021equilibrium} dived into the relationship between block rewards and the equilibrium level of staking, demonstrating that staking levels do not consistently rise with increasing block rewards. Chitra et al.~\cite{chitra2022improving} explored the impact of MEV on validators, revealing that rational validators tend to remain active by sharing a portion of the MEV revenue through redistribution, thereby maintaining economic security. John et al.~\cite{john2020economic} investigated an economic model of PoS, exploring how adoption decisions are influenced by security risks and network congestion. Gersbach~\cite{gersbach2022staking} conducted analyses on the existence and uniqueness of equilibria within staking pools, considering the presence of malicious agents, and identified potential risks. Brunjes et al.~\cite{brunjes2020reward} proposed reward-sharing schemes that encourage the fair formation of stake pools involving a large number of stakeholders. Furthermore, several studies have also focused on other PoS-based blockchain platforms, such as Cardano~\cite{brunjes2020reward} and Tezos~\cite{neuder2020defending,neuder2021selfish}.

\smallskip
\noindent\textbf{Voting power analyses.}  Mueller et al.~\cite{mueller2022understanding} employed an agent-based simulation approach to investigate the decentralization of validators' decision-making power. Messias~\cite{messias2023understanding} investigated the voting distribution and its impact on several leading DeFi protocols. Fritsch et al.\cite{fritsch2022analyzing} and Wang et al.\cite{wang2022empirical} conducted research on the distribution of voting power among governance delegates and its impact on governance decisions within blockchain DAOs~\cite{yu2023leveraging}. Li et al.~\cite{li2023how} analyzed the security issues of token-based voting governance in DPoS blockchains.

\smallskip
\noindent\textbf{Validator selection} (equiv. committee selection/formation). As discussed in \underline{Sec.\ref{sec-intro}}, most studies rely on three straightforward methods, \textit{randomized block selection} (a combination of the lowest hash value and the highest stake)~\cite{posetheruem,bsc}, \textit{coin age selection} (the duration of the tokens have been staked)~\cite{kiayias2017ouroboros,kerber2019ouroboros}, and \textit{node’s wealth} (holding shares)~\cite{king2012ppcoin}. Unfortunately, we encountered a lack of sufficient formal studies that analyze committee formation in the context of PoS. Nonetheless, we refer to several relevant materials. Gavzi et al.~\cite{gavzi2023fait} explored the trade-off between committee size and the probability of selecting a committee that may contain corrupted validators, potentially resulting in system failure.

\smallskip
\noindent\textbf{Delegation in blockchain.} Grossi~\cite{Davide2022Social} engaged in a qualitative discussion of rational delegations within liquid democracy, particularly in the context of voting for delegators. Li et al.~\cite{li2023liquid} conducted an empirical analysis of participation and delegation behaviors within DPoS-based blockchain systems. Notably, based on existing investigations, most blockchain delegation analyses have predominantly focused on \textit{permissioned} blockchains, rather than exploring the nuances of \textit{permissionless} protocols like PoS. This distinction underscores the motivation behind the present work.

\smallskip
\noindent\textbf{Game theory in blockchain.} Game theory has been widely employed as a tool for analyses in the blockchain field due to the profit-driven nature of its participants \cite{liu2019survey}.  The first aspects typically involve defining constraints, which encompass both static and dynamic aspects, or involve two-party and multi-party interactions. Once these constraints are established, a proper game theory model can be applied, such as utilizing stochastic games \cite{kiayias2016blockchain}, cooperative games \cite{lewenberg2015bitcoin}, evolutionary games \cite{kim2019mining}, and Stackelberg games \cite{chen2022absnft}. 

The second aspect entails selecting an appropriate model that aligns with the assumed conditions and applying it to real-world cases.  Lohr et al.~\cite{lohr2022formalizing} and Janin et al.~\cite{janin2020filebounty} delve into two-party exchange protocols. Analyzing deviations from the Nash equilibrium provides insights into protocol designs. Qin et al.~\cite{qin2022bdts} develop a data trading platform and conduct an analysis of the interactions among involved parties via subgame perfect Nash equilibrium.

Additionally, the game analyses can be beneficial for constructing fairness and security in mining procedures. Existing research independently examines miner strategies in contexts such as selfish mining \cite{eyal2015miner}\cite{kwon2017selfish}\cite{negy2020selfish}\cite{sapirshtein2016optimal}, multiple
miners strategy \cite{bai2021blockchain}, compliance strategies in PoW/PoS \cite{karakostas2022blockchain}, pooled mining strategies \cite{wang2019pool}\cite{li2020mining}, and fickle mining behaviors across different chains \cite{kwon2019bitcoin}.


\begin{table}[!hbt]
\caption{Notations}\label{tab:notation}
\vspace{-0.15in}
\renewcommand\arraystretch{1.1}
\begin{center}
\resizebox{1\linewidth}{!}{
\begin{tabular}{ccc} 

       \multicolumn{1}{c}{\cellcolor{gray!10} \textbf{Name}} & \cellcolor{gray!10} \textbf{Symbol} &  \cellcolor{gray!10} \textbf{Function}  \\
        \cmidrule{1-3}
        \cellcolor{gray!10} Participants &\cellcolor{gray!10}  $P$ & All participants in the market: $A\cup V$. \\
        
        \cellcolor{gray!10} User &\cellcolor{gray!10}  $A$ (or $a_i$) & The group we are focused on in this paper. \\ 

        \cellcolor{gray!10} Validator &\cellcolor{gray!10}  $V$  (or $v_j$)& The group who are eligible for producing blocks. \\ 

        \cellcolor{gray!10} Delegation &\cellcolor{gray!10}  $\textbf{d}$  (or $d_i$)& A user $i$'s choice of delegation.  \\

        \cellcolor{gray!10} Delegation &\cellcolor{gray!10}  $\textbf{t}$  (or $t_i$)& A user $i$'s choice of token number.  \\

        \cellcolor{gray!10} Delegation set &\cellcolor{gray!10}  $d(v_j)$ & The group who delegated to the validator $v_j$.\\

        \cellcolor{gray!10}Commission fee &\cellcolor{gray!10}  c & The charge of staking services, $c\in[0,1]$.  \\

         \cellcolor{gray!10} Event &\cellcolor{gray!10}  $\theta_j=1/0$  & A validator will stay in the market/game (or not). \\

         \cellcolor{gray!10} Delegation event &\cellcolor{gray!10}  $\theta_{ij}=1/0$  & User $a_i$ delegates to validator $v_j$. \\

         \cellcolor{gray!10} Integrity &\cellcolor{gray!10}  $p_j$  & \multicolumn{1}{c|}{A prior that $v_j$ stays in the he market/game.} \\
         
         \cellcolor{gray!10} Evidence quality &\cellcolor{gray!10}  $z_j$  & \multicolumn{1}{c|}{The probability that evidence $e_j$ perfectly reveal $p_j$.} \\

         \cellcolor{gray!10} Accuracy and error &\cellcolor{gray!10}  $q_{ij}, \bar{q}_{ij}$  & \multicolumn{1}{c|}{The probability that $a_i$ delegates to $v_j$ conditioned on $e_j=1$.} \\
         
         \cellcolor{gray!10} Trust &\cellcolor{gray!10}  $T$  & \multicolumn{1}{c|}{A statistic process for stimulating user's trust.} \\

         \cellcolor{gray!10} Budget &\cellcolor{gray!10}  $b_i$  & \multicolumn{1}{c|}{All tokens $a_i$ can use.} \\

         \cellcolor{gray!10} Utility &\cellcolor{gray!10}  $u_i$  & \multicolumn{1}{c|}{The utility function of $a_i$.} \\

\cmidrule{3-3}
        
\end{tabular}
}
\end{center}

\end{table}

\begin{table}[!hbt]
\caption{Mappings between \textit{providers} and \textit{PoS blockchain}s}\label{tab:staking-blockchain}
\renewcommand\arraystretch{1.1}
\begin{center}
\resizebox{1\linewidth}{!}{
\begin{tabular}{c|ccc ccc ccc ccc cc} 

           & 
         \rotatebox{75}{ \hlhref{https://lido.fi/}{Lido}} &  
         \rotatebox{75}{ \hlhref{https://www.ankr.com/staking-crypto/}{Ankr}} & 
         \rotatebox{75}{ \hlhref{https://dokia.capital/}{Dokia Capital}} & 
         \rotatebox{75}{ \hlhref{https://marinade.finance/}{Marinade Finance}} & 
         \rotatebox{75}{ \hlhref{https://p2p.org/}{P2P Validator}} & 
         \rotatebox{75}{ \hlhref{https://rocketpool.net/}{Rocket Pool}} & 
         \rotatebox{75}{ \hlhref{https://stakewise.io/}{StakeWise}} & 
         \rotatebox{75}{ \hlhref{https://www.stakewith.us/}{StakeWithUs}} & 
         \rotatebox{75}{ \hlhref{https://staked.us/}{Staked}} &
         \rotatebox{75}{ \hlhref{https://staked.us/}{Stakin}} & 
         \rotatebox{75}{ \hlhref{https://stakingfacilities.com/}{Staking Facilities}}  &
         \rotatebox{75}{ \hlhref{https://stake.fish/networks}{Stake.fish}} &
         \\
        \cmidrule{1-6} \cmidrule{11-13} 

        BNB Chain & \cellcolor{gray!10}  &   \cellcolor{gray!10} \cmark & \cellcolor{gray!10} &  & & & & & & & & \multicolumn{1}{c|}{}  \\
       
        Fantom & \cellcolor{gray!10}  &   \cellcolor{gray!10} \cmark & \cellcolor{gray!10}  &   & & & & & & & & \multicolumn{1}{c|}{}  \\
       
        Avalanche & \cellcolor{gray!10}  &   \cellcolor{gray!10} \cmark & \cellcolor{gray!10}  &  &   &   & & & & & &  \multicolumn{1}{c|}{}  \\

        Serum &  \cellcolor{gray!10} & \cellcolor{gray!10} & \cellcolor{gray!10}  \cmark &  \cellcolor{gray!10}   & \cellcolor{gray!10}   \\

       Nucypher & \cellcolor{gray!10}  & \cellcolor{gray!10}  & \cellcolor{gray!10}  &  \cellcolor{gray!10}  &  \cellcolor{gray!10} \cmark &    \\

      Regen &  \cellcolor{gray!10} &  \cellcolor{gray!10} &  \cellcolor{gray!10}  &   \cellcolor{gray!10}  &  \cellcolor{gray!10} \cmark &    \\

      DAObet & \cellcolor{gray!10}  & \cellcolor{gray!10}   & \cellcolor{gray!10}   &  \cellcolor{gray!10}  &  \cellcolor{gray!10} \cmark &   &     \\

       Marlin &  \cellcolor{gray!10}   &  \cellcolor{gray!10}   &   \cellcolor{gray!10}  &  \cellcolor{gray!10}  &  \cellcolor{gray!10} \cmark &\cellcolor{gray!10} &\cellcolor{gray!10} &\cellcolor{gray!10}   &\cellcolor{gray!10}   &\cellcolor{gray!10}     \\
       
       LSD & \cellcolor{gray!10} \cmark &  \cellcolor{gray!10} \cmark & \cellcolor{gray!10}   & \cellcolor{gray!10}   &  \cellcolor{gray!10} &  \cellcolor{gray!10}  &  \cellcolor{gray!10}   \cmark &  \cellcolor{gray!10} &\cellcolor{gray!10}  &\cellcolor{gray!10}  \\  
              
      Flow &\cellcolor{gray!10}   & \cellcolor{gray!10}  & \cellcolor{gray!10}   &  \cellcolor{gray!10}  &  \cellcolor{gray!10} \cmark &  \cellcolor{gray!10}  & \cellcolor{gray!10}  &\cellcolor{gray!10}  \cmark & \cellcolor{gray!10} &\cellcolor{gray!10}  \\

        Polygon & \cellcolor{gray!10} \cmark    &  \cellcolor{gray!10} \cmark & \cellcolor{gray!10}  & \cellcolor{gray!10}     &  \cellcolor{gray!10}\cmark  &  \cellcolor{gray!10}  &  \cellcolor{gray!10}  &  \cellcolor{gray!10}   &  \cellcolor{gray!10} \cmark &  \cellcolor{gray!10}  \cmark &  \cellcolor{gray!10}   & \cellcolor{gray!10}  \\
        
        Ethereum & \cellcolor{gray!10}  \cmark  &   \cellcolor{gray!10} \cmark & \cellcolor{gray!10}  & \cellcolor{gray!10}  \cmark  &  \cellcolor{gray!10}  \cmark &  \cellcolor{gray!10}  \cmark &  \cellcolor{gray!10}  \cmark &  \cellcolor{gray!10} \cmark  &  \cellcolor{gray!10} \cmark &  \cellcolor{gray!10}  \cmark &  \cellcolor{gray!10} \cmark  & \cellcolor{gray!10} \cmark \\
       
        Polkadot & \cellcolor{gray!10}  &  \cellcolor{gray!10} \cmark & \cellcolor{gray!10}  & \cellcolor{gray!10}   &  \cellcolor{gray!10} \cmark &  \cellcolor{gray!10}  &  \cellcolor{gray!10}  &  \cellcolor{gray!10}   &  \cellcolor{gray!10}  \cmark &  \cellcolor{gray!10}  \cmark & \cellcolor{gray!10}\cmark &  \cellcolor{gray!10} \cmark\\
       
        Kusama & \cellcolor{gray!10}  &  \cellcolor{gray!10} \cmark & \cellcolor{gray!10}  \cmark & \cellcolor{gray!10}   &  \cellcolor{gray!10} \cmark &  \cellcolor{gray!10}  &  \cellcolor{gray!10}  &  \cellcolor{gray!10}   &  \cellcolor{gray!10} \cmark &  \cellcolor{gray!10}  \cmark &  \cellcolor{gray!10}  & \cellcolor{gray!10}\cmark   \\

        Kava & \cellcolor{gray!10}  &  \cellcolor{gray!10} & \cellcolor{gray!10}  \cmark & \cellcolor{gray!10}   &  \cellcolor{gray!10} \cmark &  \cellcolor{gray!10}  &  \cellcolor{gray!10}  &  \cellcolor{gray!10} \cmark &  \cellcolor{gray!10} \cmark &  \cellcolor{gray!10}  \cmark &  \cellcolor{gray!10}  & \cellcolor{gray!10}\cmark  \\
       
        Cosmos &  \cellcolor{gray!10}  &  \cellcolor{gray!10} & \cellcolor{gray!10}  \cmark & \cellcolor{gray!10}   &  \cellcolor{gray!10} \cmark &  \cellcolor{gray!10}  &  \cellcolor{gray!10}  &  \cellcolor{gray!10} \cmark  &  \cellcolor{gray!10} \cmark &  \cellcolor{gray!10}\cmark &  \cellcolor{gray!10}\cmark  & \cellcolor{gray!10}\cmark  \\

        IRISnet & \cellcolor{gray!10}  &  \cellcolor{gray!10} & \cellcolor{gray!10}  \cmark & \cellcolor{gray!10}   &  \cellcolor{gray!10} \cmark &  \cellcolor{gray!10}  &  \cellcolor{gray!10}  &  \cellcolor{gray!10} &  \cellcolor{gray!10}  &  \cellcolor{gray!10}\cmark &  \cellcolor{gray!10}  & \cellcolor{gray!10}\cmark  \\

        Near &  \cellcolor{gray!10}  &  \cellcolor{gray!10} & \cellcolor{gray!10}  \cmark & \cellcolor{gray!10}   &  \cellcolor{gray!10} \cmark &  \cellcolor{gray!10}  &  \cellcolor{gray!10}  &  \cellcolor{gray!10}&  \cellcolor{gray!10} \cmark  &  \cellcolor{gray!10}\cmark &  \cellcolor{gray!10}  & \cellcolor{gray!10}\cmark \\
       
        Solana &  \cellcolor{gray!10} \cmark  &  \cellcolor{gray!10} & \cellcolor{gray!10}  \cmark & \cellcolor{gray!10}  \cmark  &  \cellcolor{gray!10} \cmark &  \cellcolor{gray!10}  &  \cellcolor{gray!10}  &  \cellcolor{gray!10} \cmark  &  \cellcolor{gray!10}\cmark &  \cellcolor{gray!10} \cmark  & \cellcolor{gray!10}\cmark &  \cellcolor{gray!10} \cmark\\
       
       Tezos & \cellcolor{gray!10} & \cellcolor{gray!10} & \cellcolor{gray!10}  &  \cellcolor{gray!10}  &  \cellcolor{gray!10} \cmark &  \cellcolor{gray!10}  &  \cellcolor{gray!10}   &  \cellcolor{gray!10}   &  \cellcolor{gray!10} \cmark & \cellcolor{gray!10} \cmark & \cellcolor{gray!10} \cmark &  \cellcolor{gray!10} \cmark\\

       TheGraph &  &   &  &   &  \cellcolor{gray!10} \cmark &  \cellcolor{gray!10}  &  \cellcolor{gray!10}   &  \cellcolor{gray!10}  \cmark &  \cellcolor{gray!10} \cmark &  \cellcolor{gray!10}  & \cellcolor{gray!10}  \cmark & \cellcolor{gray!10} \\
       
       Cardano &   & &   &   &  \cellcolor{gray!10} \cmark &  \cellcolor{gray!10}  &  \cellcolor{gray!10}   &  \cellcolor{gray!10}  &  \cellcolor{gray!10} \cmark &  \cellcolor{gray!10}  & \cellcolor{gray!10}  & \cellcolor{gray!10} \cmark  \\
       
       Oasis &   &   &   &    &  \cellcolor{gray!10} \cmark &  \cellcolor{gray!10}  &  \cellcolor{gray!10}   &  \cellcolor{gray!10} &  \cellcolor{gray!10}  &  \cellcolor{gray!10}  & \cellcolor{gray!10}  & \cellcolor{gray!10} \cmark \\

        Mina &   &   &   &   &  \cellcolor{gray!10} \cmark &  \cellcolor{gray!10}  &  \cellcolor{gray!10}   &  \cellcolor{gray!10}   &  \cellcolor{gray!10} \cmark &  \cellcolor{gray!10} \cmark  & \cellcolor{gray!10}  & \cellcolor{gray!10} \\

       SUI &  &  &   &   &  \cellcolor{gray!10} \cmark  &  \cellcolor{gray!10}  &  \cellcolor{gray!10}   &  \cellcolor{gray!10}  \cmark &  \cellcolor{gray!10} \cmark &  \cellcolor{gray!10} \cmark & \cellcolor{gray!10} \cmark  & \cellcolor{gray!10} \cmark \\

        APTOS &  & &  &    &  \cellcolor{gray!10} \cmark  &  \cellcolor{gray!10}  &  \cellcolor{gray!10}   &  \cellcolor{gray!10} &  \cellcolor{gray!10} \cmark  &  \cellcolor{gray!10} \cmark  & \cellcolor{gray!10} \cmark  & \cellcolor{gray!10}   \\
       
        Persistence &  &  &  &  &  \cellcolor{gray!10} \cmark &  \cellcolor{gray!10}  &  \cellcolor{gray!10}   &  \cellcolor{gray!10}  \cmark &  \cellcolor{gray!10} &  \cellcolor{gray!10}  & \cellcolor{gray!10}  & \cellcolor{gray!10} \\
       
        Osmosis &   &  &   &  &  \cellcolor{gray!10}  \cmark  &  \cellcolor{gray!10} &  \cellcolor{gray!10}   &  \cellcolor{gray!10} &  \cellcolor{gray!10}   &  \cellcolor{gray!10}  & \cellcolor{gray!10}  & \cellcolor{gray!10}\cmark \\

        JUNO &  &   &  &    & \cellcolor{gray!10} & \cellcolor{gray!10}   &   \cellcolor{gray!10}   &  \cellcolor{gray!10} \cmark  &  \cellcolor{gray!10} &  \cellcolor{gray!10}  & \cellcolor{gray!10}  & \cellcolor{gray!10}\cmark \\
       
        Band Protocol &  &  & &   &   &   &   &  \cellcolor{gray!10}  \cmark &  \cellcolor{gray!10} &  \cellcolor{gray!10}  &  \cellcolor{gray!10}  &  \cellcolor{gray!10}\cmark \\

        Loom &   &  &  &    &   &   &     &  \cellcolor{gray!10} \cmark  &  \cellcolor{gray!10} &  \cellcolor{gray!10}  & \cellcolor{gray!10}  & \cellcolor{gray!10} \cmark  \\

        SEI &   &   &    &    &     &   &   & \cellcolor{gray!10}  &  \cellcolor{gray!10} \cmark &  \cellcolor{gray!10} \cmark & \cellcolor{gray!10}   & \multicolumn{1}{c|}{\cellcolor{gray!10}} \\

        Algorand &  &  &  &   &   &   &   &   &  \cellcolor{gray!10} &  \cellcolor{gray!10} \cmark & \cellcolor{gray!10}  & \multicolumn{1}{c|}{\cellcolor{gray!10}}  \\
       
        ICON &  &  &  &   &   &   &   &     &  \cellcolor{gray!10} &  \cellcolor{gray!10} \cmark & \cellcolor{gray!10}  & \multicolumn{1}{c|}{\cellcolor{gray!10}}  \\
       
        Celo & &  &  &   &   &   &    & &  \cellcolor{gray!10}  &  \cellcolor{gray!10} \cmark & \cellcolor{gray!10}  & \multicolumn{1}{c|}{\cellcolor{gray!10}} \\

        Casper Net. &   &  &   &   &  &  &   &   &  \cellcolor{gray!10} &  \cellcolor{gray!10}  & \cellcolor{gray!10}  & \multicolumn{1}{c|}{\cellcolor{gray!10} \cmark }\\

        \cmidrule{1-3}\cmidrule{8-13}
 
\end{tabular}
}
\end{center}
\end{table}

\end{document}